\documentclass[onefignum,onetabnum]{siamonline190516}

\usepackage{amsfonts}
\usepackage{graphicx}
\usepackage{epstopdf}
\usepackage{algorithmic}
\usepackage{amssymb, kbordermatrix}
\ifpdf
  \DeclareGraphicsExtensions{.eps,.pdf,.png,.jpg}
\else
  \DeclareGraphicsExtensions{.eps}
\fi

\newsiamthm{example}{Example}

\headers{EzGP Models for QQ Factors}{Q. Xiao, A. Mandal, CD. Lin and X. Deng}

\title{EzGP: Easy-to-Interpret Gaussian Process Models for Computer Experiments with Both Quantitative and Qualitative Factors\thanks{Submitted on September 19, 2019.
}}

\author{Qian Xiao\thanks{Department of Statistics, University of Georgia, Athens, GA
  (\email{qian.xiao@uga.edu}, \email{amandal@stat.uga.edu}).}
  \and Abhyuday Mandal\footnotemark[2]
\and C. Devon Lin\thanks{Department of Mathematics and Statistics, Queen's University, Kingston, Canada
  (\email{devon.lin@queensu.ca)}.}
\and Xinwei Deng
\thanks{Department of Statistics, Virginia Tech, Blacksburg, VA
  (\email{xdeng@vt.edu}).}}

\usepackage{amsopn}

\usepackage{amsfonts}

\newcommand{\bT}{\textbf{T}}
\newcommand{\bA}{\textbf{A}}
\newcommand{\bB}{\textbf{B}}

\newcommand{\bx}{\textbf{x}}

\newcommand{\by}{\textbf{y}}

\newcommand{\bone}{\textbf{1}}

\newcommand{\bw}{\textbf{w}}
\newcommand{\bz}{\textbf{z}}

\newcommand{\bPhi}{\mbox{\boldmath${\Phi}$} }
\newcommand{\btheta}{\mbox{\boldmath${\theta}$} }
\newcommand{\bTheta}{\mbox{\boldmath${\Theta}$} }

\newcommand{\bm}[1]{\mbox{\boldmath$ #1 $\unboldmath}}

\def\TT{{\mbox{\tiny T}}}

\newcommand{\rb}[1]{{ #1}}

\newcommand{\nx}[1]{{{#1}}}

\ifpdf
\hypersetup{
  pdftitle={EzGP: Easy-to-Interpret Gaussian Process Models for Computer Experiments with Both Quantitative and Qualitative Factors},
  pdfauthor={Q. Xiao, A. Mandal, CD. Lin and X. Deng}
}
\fi

\begin{document}

\maketitle

\begin{abstract}
Computer experiments with both quantitative and qualitative (QQ) inputs are commonly used in science and engineering applications.
Constructing desirable emulators for such computer experiments remains a challenging problem.
In this article, we propose an easy-to-interpret Gaussian process (EzGP) model for computer experiments to reflect the change of the computer model under the different level combinations of qualitative factors.
The proposed modeling strategy, based on an additive Gaussian process, is flexible to address the heterogeneity of computer models involving multiple  qualitative factors.
We also develop two useful variants of the EzGP model to achieve computational efficiency for data with high dimensionality and large sizes.
The merits of these models are illustrated by several numerical examples and a real data application.
\end{abstract}



\section{Introduction}
\label{intro}
Computer experiments are now ubiquitous in scientific researches and engineering.
The computer models used in computer experiments are often very complex and computationally expensive, and thus require emulators in the analysis  \cite{fang2005}.
Gaussian process (GP) models, a.k.a. Kriging, have been used as a core tool for modeling computer
experiments \cite{fang2005, santner2003}. The conventional GP models often only consider quantitative inputs; while many practical applications have both quantitative and qualitative (QQ) inputs, e.g., the data center computer experiment  \cite{qian2008}, the epidemiology study \cite{bhuiyan2014}, the bio-engineering computer experiment  \cite{han2009}, \nx{the study of high performance computing systems \cite{zhang2020mixed}}, and the finite element modeling of full-scale embankment \cite{deng2017additive, rowe2015}.

For emulating computer experiments with qualitative factors, a naive approach would conduct distinct GP models for data collected at the different level combinations of the qualitative factors.
Clearly, such an approach is unwise as there could be many level combinations of the qualitative factors, and it could overlook possible dependency between responses (or outputs) at the different level combinations of the qualitative factors \cite{qian2008}.
Alternatively, it would be natural to consider the use of indicator variables, often applied in linear models, to address GP models with qualitative factors.
\rb{However, a counter example is given to show that using indicator variables for qualitative factors in the multiplicative correlation function is problematic \cite{zhang2015computer}.} We will further illustrate this problem in \cref{eg3} of \cref{GF}.
In this work, we propose an easy-to-interpret Gaussian process (EzGP) model to appropriately use indicator functions in additive GP models for incorporating qualitative factors with meaningful interpretations and accurate predictions.

\begin{figure}[ht]
\caption{\it The response curve with respect to the quantitative factor $x$ under each level combination of the two qualitative factors}
  \begin{center}
  \begin{tabular}{cccc}
    \includegraphics[width=0.21\linewidth]{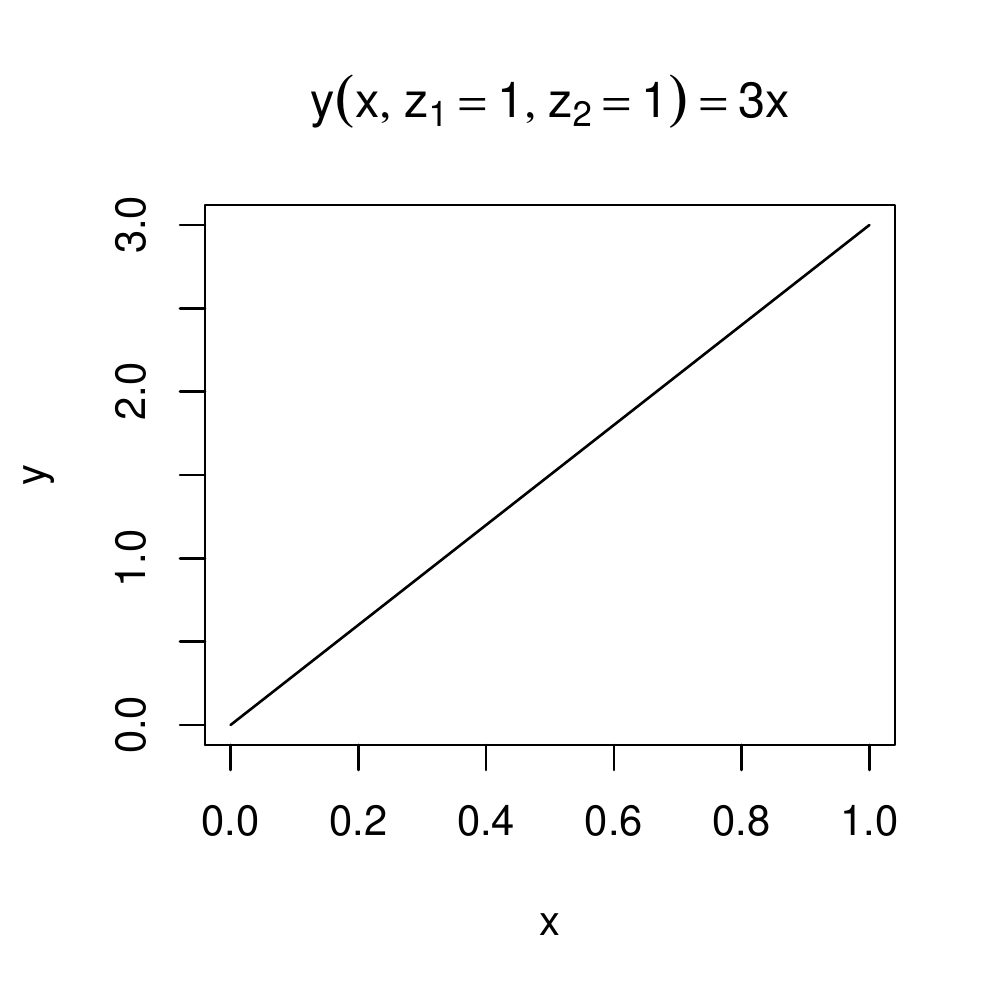} &
    \includegraphics[width=0.21\linewidth]{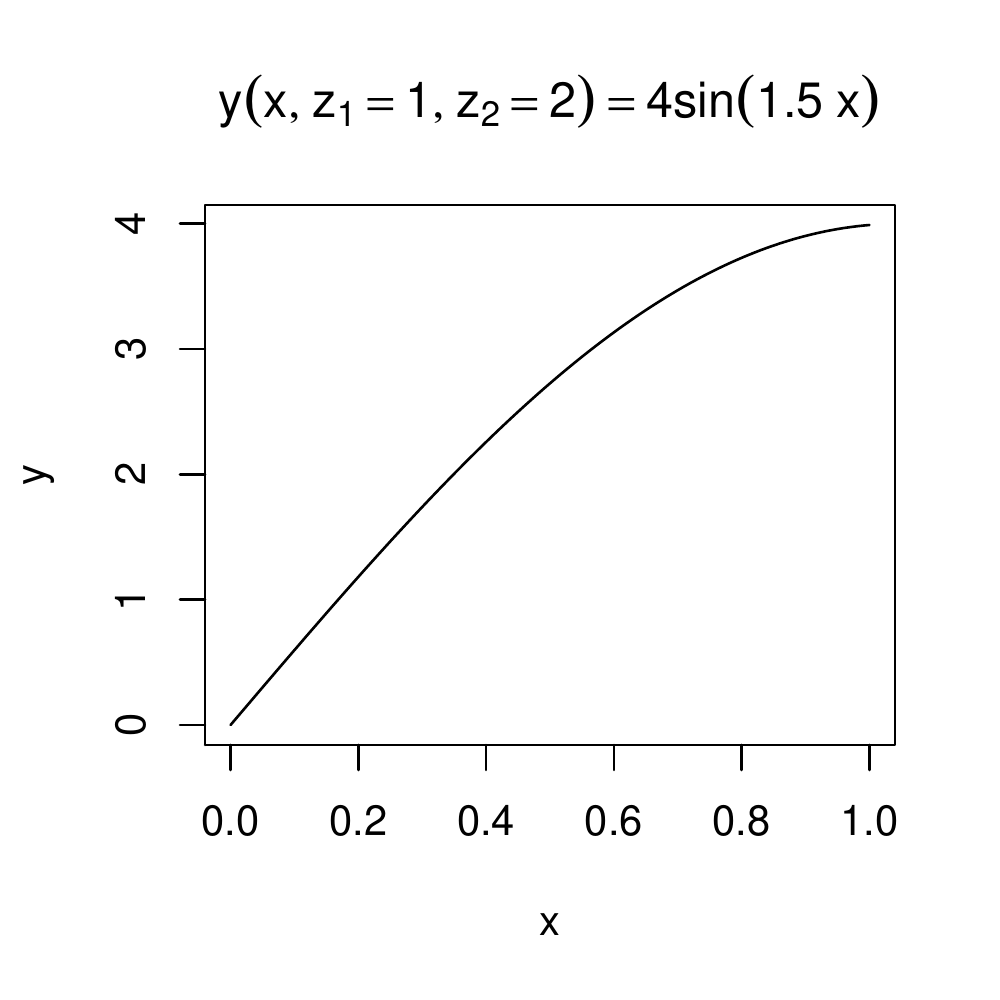} &
    \includegraphics[width=0.21\linewidth]{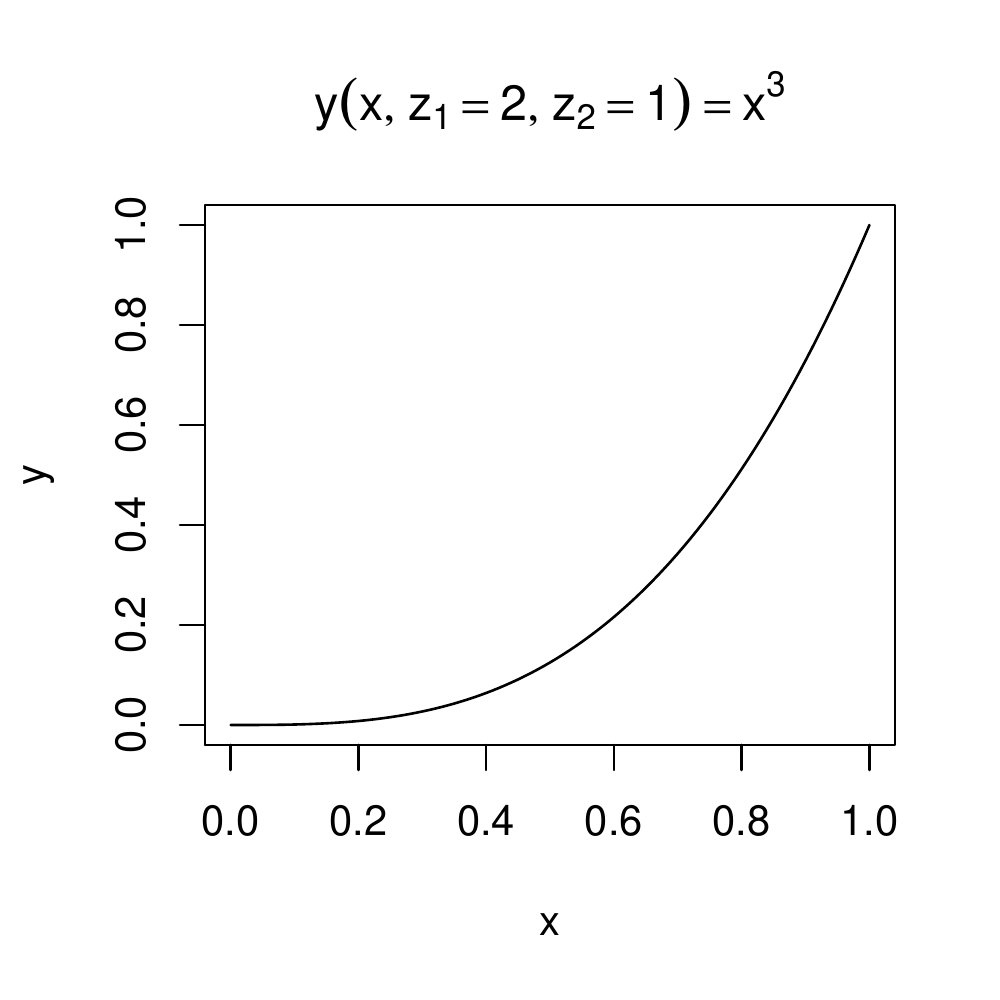} &
    \includegraphics[width=0.21\linewidth]{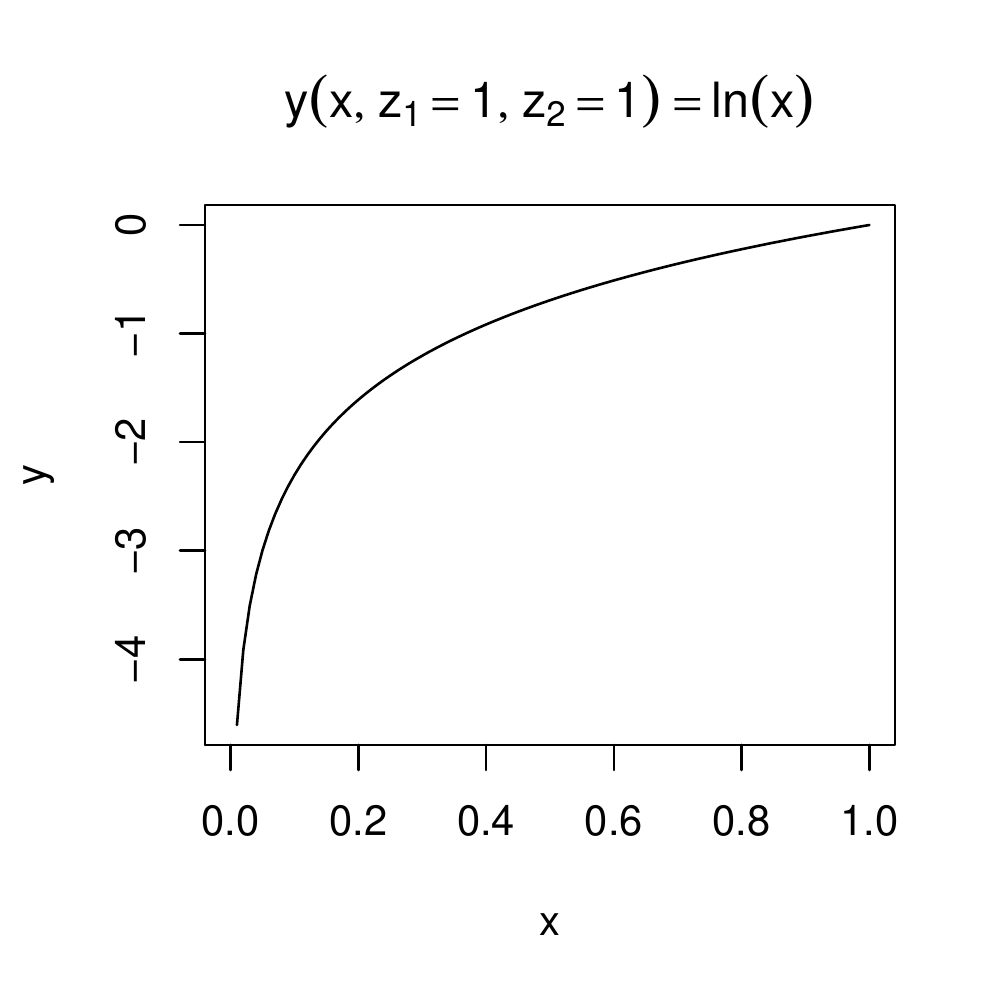} \\
  \end{tabular}
  \end{center}
    \label{fig:intro}
\end{figure}

For GP models of computer experiments with both QQ inputs, many existing works focus on  constructing correlations between the levels for each qualitative factor, and then use the multiplicative structure to link them with the correlation functions for quantitative factors \cite{qian2008, zhou2011}. Such a multiplicative correlation function requires the ``shape" of local variation as a function of quantitative factors to be the same for all level combinations of the qualitative factors; that is, the correlation parameters and process variances are the same for different qualitative level combinations \cite{zhang2015computer}.
This is a strong assumption since computer models can be quite different for distinct qualitative level combinations, especially when there are multiple qualitative factors.
\rb{Such a way to construct correlation functions for qualitative factors is also applied to the additive GP models in \cite{deng2017additive}. Yet, it may not be interpretable in practice.
As an illustration,} consider a computer experiment with one quantitative factor $x$ and two qualitative factors $z_1$ and $z_2$ each having two nominal values.
For its four different qualitative level combinations, the corresponding computer models are $3x$, $4\sin(1.5x)$,  $x^3$ and $\text{ln}(x)$, which are shown in \cref{fig:intro}.
Here, it is not easy to interpret if one simply uses a scalar value, i.e., the correlation between two levels for each qualitative factor, to quantify the complex relationship between different functions of computer models \cite{deng2017additive,qian2008}.
\rb{It would be more natural to use indicator functions to reflect the GP being adjusted from a base GP under the different level combinations of the qualitative factors.}

\rb{In this article, we first lay out a general additive GP structure,
and then make several reasonable assumptions to appropriately adopt indicator functions in developing the proposed EzGP model.}
The proposed method has a clear interpretation of its additive covariance structure to reflect the relationship between the response and the quantitative factors and identify how the qualitative factors affect such relations.
It is suitable for dealing with discontinuities in response surfaces due to qualitative factors in computer experiments.
The key idea of the proposed EzGP model is to start with a base GP accounting for only quantitative factors, and have GP components in an additive fashion to adjust the different level combinations of the qualitative factors.
It follows a similar spirit of using indicator functions in variance decomposition, but at the scope of Gaussian process under each level of the qualitative factors.
Compared to existing models using scalars to quantify the correlations between levels in the qualitative factors  \cite{deng2017additive, qian2008}, 
the EzGP model does not explicitly construct the correlation functions for the qualitative factors. Instead, 
it quantifies the relationships among different response surfaces under the different level combinations of the qualitative factors 
through an additive combination of several GPs, which leads to an easy-to-interpret covariance structure.

\rb{The EzGP model is proposed for computer experiments with QQ inputs where \nx{multiple qualitative factors} are involved.
Specifically, we \nx{focus on} complex computer experiments where differences between the computer models for the distinct level combinations are \nx{large}.
In such cases, the types of functions in computer models can be different, e.g. the toy example in  \cref{fig:intro}, the simulation example in \cite{zhang2015computer} and some real data in \cite{du2017, rowe2015}.}
Based on the EzGP model, we further develop a variant, the efficient EzGP (EEzGP) model, suitable for computer experiments with a large number of qualitative factors.
We also develop another variant, the localized EzGP (LEzGP) method, to efficiently deal with large sample sizes.

The remainder of this article is organized as follows.
\Cref{nb} provides a brief introduction to GP models and review existing methods.
\Cref{gagps} details the proposed EzGP, EEzGP and LEzGP methods.
\Cref{ss}  presents several numerical examples and \cref{re} reports a real application of the proposed models.
\Cref{cd} concludes this work and discusses some future work.
All proofs and technical details are relegated to the Appendix.

\section{Notation and Literature Review}
\label{nb}

In this section, we introduce notation and review some current literature.
Throughout this paper, we consider an $n$-run  computer experiment with $p$ quantitative factors and $q$ qualitative factors. We denote the $i^{th}$ quantitative factor as $x^{(i)}$ ($i=1, \ldots, p$) and the $j^{th}$ qualitative factor as $z^{(j)}$ ($j=1, \ldots, q$). There are $m_j$ levels ($\{1, \ldots, m_j\}$) of the qualitative factor $z^{(j)}$. Denote the $k^{th}$ ($k = 1, \ldots, n$) data input as $\bw_{k} = (\bx_{k}^\TT, \bz_{k}^\TT)^\TT$ where $\bx_k = (x_{k1},\ldots,x_{kp})^\TT \in \mathbb{R}^{p}$ is the quantitative part and $\bz_k = (z_{k1},\ldots,z_{kq})^\TT \in \mathbb{N}^{q}$ is the qualitative part (coded in levels) of the input. Denote $Y(\bw_{k})$ as the output from the input $\bw_{k}$ and the response (or output) vector $\by = (Y(\bw_{1}), \ldots, Y(\bw_{n}))^\TT$.

In the standard GP model \cite{kleijnen2009, rasmussen2006,sacks1989}, the inputs are all quantitative and the outputs can be  viewed as realizations of a GP. The correlation
between outputs is determined by a stationary correlation function, e.g.,  Gaussian, power-exponential and Mat\'ern correlation functions. To model the relationship between outputs $Y(\bx)$ and inputs $\bx$, one popular GP model, known as an ordinary GP (Kriging) model, assumes,
\begin{equation}\label{eq:gp2}
Y(\bx) = \mu  + G(\bx),
\end{equation}
\noindent
where $\mu$ is the constant mean,
$G(\bx)$ is a GP with zero mean and the covariance function $\phi(\cdot) = \sigma^2 R(\cdot|\btheta)$.
A popular choice for $R(\cdot|\btheta)$ is the Gaussian correlation function
\begin{equation}
\label{gc}
R(\bx_{i},\bx_{j}|\btheta) = \hbox{exp} \{ - \sum_{k=1}^p \theta_k  (x_{ik} - x_{jk})^2\},
\end{equation}
\noindent where two inputs $\bx_{i} = (x_{i1}, \ldots, x_{ip})^\TT$ and $\bx_{j} = (x_{j1}, \ldots, x_{jp})^\TT$, and the correlation parameters $\btheta=(\theta_1,\ldots,\theta_p)^\TT$ with all $\theta_k > 0$ ($k =1, \ldots, p$).

To deal with QQ inputs, a popular GP based model \cite{qian2008, zhou2011} was introduced among many others \cite{han2009, swiler2014surrogate,zhang2015computer,zhang2018latent}.
Specifically, an ordinary GP model with a multiplicative covariance function is considered (for any two inputs $\bw_1$ and $\bw_2$):
\begin{equation}
\label{corm1}
\hbox{Cov}(Y(\bw_1), Y(\bw_2)) =  \sigma^2  \prod_{j=1}^{q}\tau^{(j)}_{z_{1j}z_{2j}} R(\bx_{1},\bx_{2}|\btheta) ,
\end{equation}
where the parameter $\tau^{(j)}_{z_{1j}z_{2j}}$ represents the correlation between two levels ($z_{1j}$ and $z_{2j}$) in the $j^{th}$ qualitative factor $z^{(j)}$, and $R(\bx_{1},\bx_{2}|\btheta)$ is defined in \cref{gc}.  Denote $\bT_j = (\tau^{(j)}_{z_{1j} z_{2j}})_{m_j \times m_j}$ as the correlation matrix for $z^{(j)}$ ($j = 1, \ldots, q$). Three different functions of $\tau^{(j)}_{z_{1j} z_{2j}}$ can be used:
\begin{enumerate}
\item the exchangeable correlation function (EC) \cite{joseph2007}: $\tau^{(j)}_{z_{1j} z_{2j}}= c$ ($0 < c < 1$) when $z_{1j} \neq z_{2j}$; otherwise, $\tau^{(j)}_{z_{1j} z_{2j}}= 1$ ;

\item the multiplicative correlation function (MC) \cite{mcmillan1999}: $\tau^{(j)}_{z_{1j} z_{2j}}= \hbox{exp}\{-(\theta_{z_{1j}} + \theta_{z_{2j}})\}$  when $z_{1j} \neq z_{2j}$; otherwise, $\tau^{(j)}_{z_{1j} z_{2j}}= 1$, where $\theta_{z_{1j}}, \theta_{z_{2j}} >0$;

\item the unrestrictive correlation function (UC)  \cite{qian2008, zhou2011}: define $T_j = L_jL_j^\TT$ where $L_j$ is a lower triangular matrix; for the $r^{th}$ row ($l_{r1}^{(j)}, \ldots, l_{rr}^{(j)}$) in $L_j$, $l^{(j)}_{11} = 1$, and for $r = 2, \ldots, m_j$,
\[
    \left\{
    \begin{array}{ll}
     l^{(j)}_{r1} = \cos(\varphi_{j,r,1}) \nonumber \\
l^{(j)}_{rs} = \sin(\varphi_{j,r,1}) \ldots \sin(\varphi_{j,r,s-1}) \cos(\varphi_{j,r,s}), \ \  \hbox{for} \ s = 2, \ldots, r-1 \nonumber \\
l^{(j)}_{rr} = \sin(\varphi_{j,r,1}) \ldots \sin(\varphi_{j,r,r-1}), \nonumber
    \end{array}
              \right.
  \]
where $\varphi_{j,r,s} \in (0,\pi)$ for $s = 1, \ldots, r-1$.
\end{enumerate}
We denote these three multiplicative GP models as the EC, MC and UC models, respectively.

An additive GP model was proposed in  \cite{deng2017additive}, which adopts an additive covariance function:
\begin{equation}
\label{dc}
\hbox{Cov}(Y(\bw_1), Y(\bw_2)) = \sum_{j=1}^{q}\sigma_j^2\tau^{(j)}_{z_{1j}z_{2j}}R(\bx_1, \bx_2 \vert \btheta^{(j)})
\end{equation}
where $\sigma_j^2$ and $\btheta^{(j)}$ ($j = 1, \ldots, q$) are the process variance parameters and the correlation parameters corresponding to $z^{(j)}$, respectively. The same as above, three different choices of $\tau^{(j)}_{z_{1j}z_{2j}}$: the exchangeable, multiplicative and unrestrictive correlation functions, can be adopted in \cref{dc}. We denote them as the AD\_EC, AD\_MC and AD\_UC models, respectively.
Note that if any $\tau^{(j)}_{z_{1j} z_{2j}}$ has a zero (or near zero) value, the overall covariance in \cref{corm1} will be zero (or near zero).
Such problems are avoided in the additive model structure in \cref{dc}.

\section{Easy-to-Interpret Gaussian Process (EzGP) Models}
\label{gagps}
In this section, we first lay out a general additive GP structure and describe in details the proposed EzGP model.
Then we illustrate the Efficient EzGP (EEzGP) model for data with   many qualitative factors, and discuss the Localized EzGP (LEzGP) method for data of large run sizes.

\subsection{The EzGP Model} \label{GF}
For an $n$-run computer experiment with $p$ quantitative factors and $q$ qualitative factors, we model the output at $\bw = (\bx^\TT, \bz^\TT)^\TT$ as
\begin{equation}\label{eq:additiveGP}
Y(\bw) = \mu + G_{\bz}(\bx).
\end{equation}
\rb{It means that for any level combination of $\bm z$, $Y(\bw)$ is a Gaussian process.
Specifically, we consider
\begin{equation}
\label{eq:agp}
G_{\bz}(\bx) = G_{0}(\bx)+G_{z^{(1)}}(\bx) + \cdots + G_{z^{(q)}}(\bx),
\end{equation}
where $G_0$ and $G_{z^{(h)}}$ ($h = 1, \ldots q$) are independent Gaussian processes with mean zero and the covariance functions $\phi_0$ and $\phi_h$ ($h = 1, \ldots q$), respectively.}
Here, $G_{0}$ is a standard GP taking only quantitative inputs $\bx$, which can be viewed as the base GP reflecting the intrinsic relation between $y$ and $\bx$.
The standard Gaussian covariance function is adopted for $G_{0}$, which is
\begin{equation}\label{eq:cor}
\phi_0(\bx_{i},\bx_{j}|\btheta_{0}) = \sigma_{0}^{2} \hbox{exp} \left\{ - \sum_{k=1}^p \theta_{k}^{(0)}  (x_{ik} - x_{jk})^2 \right\},
\end{equation}
where the correlation parameters in $\btheta_0=(\theta_1^{(0)},\ldots,\theta_p^{(0)})^\TT$ are all positive.

\rb{The $G_{z^{(h)}}$ can be viewed as an adjustment to the base GP by the impact of the qualitative factor $z^{(h)}$ ($h = 1, \ldots q$).}
It is a GP component concerning the $h^{th}$ qualitative factor coupled with all quantitative factors.
A general covariance function could be
\begin{align}
\label{gcf}
\phi_h((\bx_{i}^\TT, z_{ih})^\TT, (\bx_{j}^\TT, z_{jh})^\TT) =
\sigma_h^2 \hbox{exp} \left\{ - \sum\limits_{k=1}^p \theta^{(h)}_{k z_{ih} z_{jh}} (x_{ik} - x_{jk})^2 \right\} \tau^{(h)}_{z_{ih} z_{jh}},
\end{align}
where the correlation parameters $\theta^{(h)}_{k z_{ih} z_{jh}}$ ($k = 1, \ldots, p$) are specific to the  pair of levels ($z_{ih},z_{jh}$) in $z^{(h)}$, and $\tau^{(h)}_{z_{ih} z_{jh}}$ and $\sigma_h^2$ are defined in \cref{corm1} and \cref{dc}, respectively.
Clearly, such a general form involves too many parameters and thus is hard to interpret.
Note that the additive model in \cite{deng2017additive} can be viewed as a special case of the general model in \cref{eq:additiveGP} when simplifying all $\theta^{(h)}_{k z_{ih} z_{jh}} = \theta^{(h)}_{k}$ for any ($z_{ih},z_{jh}$) in \cref{gcf} and not considering the base $G_0$ in \cref{eq:agp}. As discussed in \cref{intro}, such a simplification may not be reasonable in some practical cases. Below, we will introduce the EzGP model which simplifies \cref{gcf} in a more meaningful way. When there are at least two qualitative factors, its structural formulation will be different from the additive model in \cite{deng2017additive} regardless of the choice for $\tau^{(h)}_{z_{ih} z_{jh}}$.

{For the formulation in \cref{eq:agp}, the base GP $G_0$ is adjusted by GP component $G_{z^{(h)}}$ to account for the effect of different levels in $z^{(h)}$.
This is analogous to using  indicator functions in variance decomposition.
To enable an easy-to-interpret model, we consider the covariance function of $G_{z^{(h)}}, h = 1, \ldots, q$ as}
\begin{align}\label{eq: cor-1}
\phi_h((\bx_{i}^\TT, z_{ih})^\TT, (\bx_{j}^\TT, z_{jh})^\TT | \bTheta^{(h)}) =
\sigma_h^2 \hbox{exp} \left\{ - \sum\limits_{k=1}^p \theta^{(h)}_{kl_h}  (x_{ik} - x_{jk})^2 \right\} I(z_{ih} = z_{jh} \equiv l_h),
\end{align}
where $z_{ih}$ and $z_{jh}$ are the levels of $z^{(h)}$ in the $i^{th}$ and $j^{th}$ inputs, respectively; $\sigma_h^2$ is the variance parameter for $z^{(h)}$; $l_h$ takes values in \{$1, \ldots, m_h$\} and $m_h$ is the number of levels in $z^{(h)}$; $\bTheta^{(h)} = (\theta^{(h)}_{kl_h})_{p \times m_{h}}$ is the matrix for correlation parameters; the indicator function $I(z_{ih} = z_{jh} \equiv l_{h}) = 1$ for $z_{ih} = z_{jh} \equiv l_{h}$, otherwise 0. Without any prior information, we can assume that different levels in $z^{(h)}$ will result in different and  independent Gaussian processes, and thus  $\phi_h((\bx_{i}^\TT, z_{ih})^\TT, (\bx_{j}^\TT, z_{jh})^\TT | \bTheta^{(h)}) = 0$ when $z_{ih} \neq z_{jh}$. For distinct levels $l_h$ in $z^{(h)}$, parameters $\theta^{(h)}_{kl_h}$ are different, thus we have different Gaussian processes to depict different computer models associated with the different levels of the qualitative factors. Such a strategy makes the GP model structure parsimonious and easy to interpret, which avoids directly modeling the correlation functions of qualitative factors.

Based on \cref{eq:cor,eq: cor-1,eq:agp}, for any two inputs $\bw_i$ and $\bw_j$, the covariance function for the model in  \cref{eq:additiveGP} can be specified by
\begin{align}
\label{eq:corr-fun}
\phi(\bw_i, \bw_j) & =  \hbox{Cov} ( Y(\bw_i), Y(\bw_j) ) \nonumber\\
& =  \phi_0(\bx_{i},\bx_{j}|\btheta_{0}) + \sum_{h  =1}^{q} \phi_h(\bx_{i},\bx_{j}|\bTheta^{(h)}) \nonumber \\
& = \sigma_{0}^{2}\hbox{exp} \{ - \sum_{k=1}^p \theta_{k}^{(0)}  (x_{ik} - x_{jk})^2\} \nonumber \\
&   + \sum_{h=1}^{q}  \sum_{l_{h}=1}^{m_{h}} I(z_{ih} = z_{jh} \equiv l_{h}) \sigma_{h}^{2} \hbox{exp} \left\{ - \sum_{k=1}^p \theta^{(h)}_{kl_{h}}  (x_{ik} - x_{jk})^2\right\}.
\end{align}
This aggregated covariance function has $(1 + p + q + p\sum_{h=1}^{q}m_h)$ parameters  which are estimated simultaneously via the maximum likelihood estimation. The following example illustrates the formulation of the EzGP model and its implication.
\begin{example}
\label{eg1}
Consider a computer experiment with two quantitative factors $x^{(1)}$ and $x^{(2)}$, and two qualitative factors $z^{(1)}$ and $z^{(2)}$ each having two levels.
Suppose that three inputs are
$\bw_1 = (x_{11} = a, x_{12} = b, z_{11} = 1, z_{12} = 2)^\TT$,
$\bw_2 = (x_{21} = c, x_{22} = d, z_{21} = 1, z_{22} = 2)^\TT$ and
$\bw_3 = (x_{31} = c, x_{32} = d, z_{31} = 2, z_{32} = 1)^\TT$
where $a$, $b$, $c$ and $d$ are arbitrary real numbers.
Here $x_{ij}$ and $z_{ij}$ represent the $j^{th}$ entry in $\bx_i$ and $\bz_i$, respectively. According to the covariance function in \cref{eq:corr-fun}, we have
\begin{align*}
& \phi(\bw_1, \bw_2) \\
& =  \sigma_0^{2}\hbox{exp} \left\{ - \sum_{k=1}^2 \theta_{k}^{(0)}  (x_{1k} - x_{2k})^2\right\}  + \sum_{l_1=1}^{2} \sigma_1^2 \hbox{exp} \left\{ - \sum_{k=1}^2 \theta^{(1)}_{kl_1} (x_{1k} - x_{2k})^2 \right\}  \\
&   \times I(z_{11} = z_{21} \equiv l_1) + \sum_{l_2=1}^{2} \sigma_2^2 \hbox{exp} \left\{ - \sum_{k=1}^2 \theta^{(2)}_{kl_2}  (x_{1k} - x_{2k})^2 \right\} I(z_{12} = z_{22} \equiv l_2) \\
& =  \sigma_0^{2}\hbox{exp} \left\{ - \theta_{1}^{(0)}  (a - c)^2 - \theta_{2}^{(0)}  (b - d)^2 \right\} + \sigma_1^2 \hbox{exp} \left\{ - \theta^{(1)}_{11}  (a - c)^2 - \theta^{(1)}_{21}  (b - d)^2 \right\} \\
&   + \sigma_2^2 \hbox{exp} \left\{ - \theta^{(2)}_{12}  (a - c)^2 - \theta^{(2)}_{22}  (b - d)^2 \right\}.
\end{align*}
Similarly, we have
\begin{equation*}
\phi(\bw_1, \bw_3) = \sigma_0^{2}\hbox{exp} \left\{ - \theta_{1}^{(0)}  (a - c)^2 - \theta_{2}^{(0)}  (b - d)^2 \right\}.
\end{equation*}
Clearly, $\hbox{Cov}(Y(\bw_1), Y(\bw_2)) \geqslant \hbox{Cov}(Y(\bw_1), Y(\bw_3))$.
In the EzGP model, it is straightforward to derive that all variances are equal; that is, $\hbox{Var}(\bw_i) = \sum_{i=0}^{q}\sigma_i^2$. Thus, we have $\hbox{Cor}(Y(\bw_1), Y(\bw_2)) \geqslant  \hbox{Cor}(Y(\bw_1), Y(\bw_3))$, which is meaningful for interpretation.
Given the inputs $\bw_2$ and $\bw_3$ having the same quantitative part, it is straightforward that $(\bw_1, \bw_2)$ should be more similar compared to $(\bw_1, \bw_3)$, since $\bw_1$ and $\bw_2$ have the same qualitative part but $\bw_1$ and $\bw_3$ do not. Thus, the correlation between $(\bw_1, \bw_2)$ should be larger than that between $(\bw_1, \bw_3)$, \rb{when at least one of the qualitative factor is significant (i.e., one of the $\sigma_1^2$ and $\sigma_2^2$ is not 0)}.
Refer to \cref{eg3} for an opposite case.
\end{example}

\begin{lemma}
\label{mezgp}
Let $\bA_0 = (\sigma_0^2 R(\bx_i, \bx_j \vert \btheta_0))_{n \times n}$ and $\bA_{hl_h} = (\sigma_h^2 R(\bx_i, \bx_j \vert \btheta_{l_h}^{(h)}))_{n \times n}$ where  $R(\cdot|\btheta)$ is defined in \cref{gc} and  $\btheta_{l_h}^{(h)} = (\theta^{(h)}_{kl_h})_{p \times 1}$.
The covariance matrix of the output vector $\bm y$ induced by the covariance function in \cref{eq:corr-fun} can be written as
\begin{equation}\label{eq in lmm1}
\hbox{Cov}(\bm y) = ( \phi(\bw_i, \bw_j) )_{n\times n} = \bA_0 + \sum_{h=1}^{q} \sum_{l_h=1}^{m_h} (\bB_{hl_h}\bB_{hl_h}^\TT)  \circ \textbf{A}_{hl_h},
\end{equation}
where $\circ$ is the Schur product (or Hadamard product). Here $\bB_{hl_h} = \textbf{E}_{h}(\textbf{I}_{m_h})_{l_h}$ where $(\textbf{I}_{m_h})_{l_h}$ is the $l_h^{th}$ column of the identity matrix $\textbf{I}_{m_h}$, $m_h$ is the number of levels in $z^{(h)}$, and $\textbf{E}_h$ is an $n \times  m_h$ expansion matrix of which each row is the dummy coding for the corresponding level in $z^{(h)}$.
\end{lemma}

\begin{example}
\rb{To illustrate matrices $\bB_{hl_h}$ and $\textbf{E}_h$ in  \cref{mezgp}, consider the column for $z^{(1)}$ in a computer experiment with 4 runs. In $\textbf{E}_1$, we use dummy coding $(1,0,0)$, $(0,1,0)$ and $(0,0,1)$ to code levels 1, 2 and 3, respectively. For $h=1$ and $l_1 = 2$, we have}
\[ z^{(1)} =
\begin{bmatrix}
1 \\
2 \\
3 \\
2 \\
\end{bmatrix},
\textbf{B}_{12} = \textbf{E}_{1}(\textbf{I}_{3})_{2}
=
\begin{bmatrix}
1&0 &0 \\
0 & 1 & 0 \\
0&0 &1 \\
0&1 & 0 \\
\end{bmatrix}
\begin{bmatrix}
0 \\
1 \\
0 \\
\end{bmatrix}
=
\begin{bmatrix}
0 \\
1 \\
0 \\
1 \\
\end{bmatrix},
\textbf{B}_{12}\textbf{B}_{12}^\TT
=
\begin{bmatrix}
0 & 0 & 0 & 0 \\
0 & 1 & 0 & 1 \\
0 & 0 & 0 & 0 \\
0 & 1 & 0 & 1 \\
\end{bmatrix}.
\]
 \end{example}

\cref{mezgp} provides insights on the covariance structure of the EzGP model. In \cref{eq in lmm1}, matrix $\bA_0$ serves as the base which corresponds to all quantitative inputs, matrix $\bB_{hl_h}\bB_{hl_h}^\TT$ selects all pairs of data that satisfy $z_{ih} = z_{jh} \equiv l_{h}$, and matrix $\textbf{A}_{hl_h}$ measures the adjustment due to the level $l_h$ in qualitative factor $z^{(h)}$ ($h = 1, \ldots, q$). Based on \cref{mezgp}, we can prove the following \cref{pd}.

\begin{lemma} \label{pd}
Given $n$ inputs $\bw_i = (\bx_i^\TT, \bz_i^\TT)^\TT$ ($i=1,\ldots,n$), the covariance matrix of the output vector $\bm y = (Y(\bw_1), \ldots, Y(\bw_n))^\TT$ induced by the covariance function in \cref{eq:corr-fun} is positive semi-definite, i.e., $\hbox{Cov}(\bm y)$ in \cref{eq in lmm1} is positive semi-definite.
\end{lemma}
\cref{pd} holds for any $\bw_1, \ldots, \bw_n$, including duplicated inputs.
For appropriate model inference, $\hbox{Cov}(\bm y)$ needs to be positive definite, and the following \cref{pdc} and  \cref{pdcc} shed some lights on this aspect.

\begin{lemma}
\label{pdc}
Given $n$ inputs $\bw_i = (\bx_i^\TT, \bz_i^\TT)^\TT$ ($i=1,\ldots,n$), if there exists an $h$ ($h = 1, \ldots, q$) such that any two inputs $\bw_i$ and $\bw_j$ ($i \neq j$) have distinct quantitative parts ($\bx_i \neq \bx_j$) whenever they have the same level in $z^{(h)}$, the covariance matrix $\hbox{Cov} (\bm y)$ induced by the covariance function in \cref{eq:corr-fun} is positive definite.
\end{lemma}

\begin{corollary}
\label{pdcc}
If there are no duplicated runs in the quantitative part of the design matrix, that is $\bx_i \neq \bx_j$ for $i \neq j$, the covariance matrix $\hbox{Cov}(\bm y)$ induced by the covariance function in \cref{eq:corr-fun} is positive definite.
\end{corollary}

\cref{pdcc} is a special case of  \cref{pdc},  and its assumption is standard in computer experiments.
If Latin hypercube designs or space-filling designs \cite{dean2015handbook} are used for quantitative factors, it is clear that $\hbox{Cov}(\bm y)$ is positive definite by \cref{pdcc}.
When the conditions in \cref{pdc} are not satisfied, one can simply add a nugget term to make the covariance matrix positive definite, which is a standard technique in Kriging  \cite{ kleijnen2009, rasmussen2006, ranjan2011computationally}.

Besides the additive covariance function in \cref{eq:corr-fun}, one could think of using indicator functions under a multiplicative covariance structure as
\begin{align}\label{eq:nested-corr-fun}
 \phi^{*}(\bw_i, \bw_j) \nonumber &= \hbox{Cov} ( Y(\bw_i), Y(\bw_j) ) = \sigma^{2} R(\bx_{i},\bx_{j}|\btheta_{0}) \prod_{h=1}^{q}R(\bx_{i},\bx_{j}|\bTheta^{(h)}) \nonumber \\
&= \sigma^{2}\hbox{exp} \left\{ - \sum_{k=1}^p \theta_{k}^{(0)}  (x_{ik} - x_{jk})^2 \right\} \nonumber \\
&  \times \prod_{h=1}^{q} \prod_{l_h=1}^{m_h} \left[ \hbox{exp} \left\{ - \sum_{k=1}^p \theta^{(h)}_{kl_{h}}  (x_{ik} - x_{jk})^2 \right\} \right] ^{I(z_{ih} = z_{jh} \equiv l_{h})}.
\end{align}
However, such a covariance function may not properly quantify the correlation for two inputs as illustrated in the following example.

\begin{example}
\label{eg3}
For the three inputs $\bw_1$, $\bw_2$ and $\bw_3 $ in \cref{eg1}, under the multiplicative covariance function in  \cref{eq:nested-corr-fun}, we have:
\begin{align*}
\phi^{*}(\bw_1, \bw_2) & =  \sigma^{2}\hbox{exp} \left\{ - \sum_{k=1}^2 \theta_{k}^{(0)}  (x_{1k} - x_{2k})^2\right\} \hbox{exp} \left\{ - \sum_{k=1}^2 \theta^{(1)}_{k1}  (x_{1k} - x_{2k})^2\right\} \\
&   \times \hbox{exp} \left\{ - \sum_{k=1}^2 \theta^{(2)}_{k2}  (x_{1k} - x_{2k})^2\right\} \\
& =  \sigma^{2}\hbox{exp} \{ - (\theta_{1}^{(0)} + \theta^{(1)}_{11} + \theta^{(2)}_{12})  (a - c)^2 - (\theta_{2}^{(0)} + \theta^{(1)}_{21} + \theta^{(2)}_{22})(b - d)^2 \},
\end{align*}
\begin{equation*}
\phi^{*}(\bw_1, \bw_3) = \sigma^{2}\hbox{exp} \Big\{ - \sum_{k=1}^2 \theta_{k}^{(0)}  (x_{1k} - x_{2k})^2 \Big\} =  \sigma^{2}\hbox{exp} \big\{ - \theta_{1}^{(0)}  (a - c)^2 - \theta_{2}^{(0)}(b - d)^2 \big\}.
\end{equation*}
It is easy to derive that $\hbox{Cor}(Y(\bw_1), Y(\bw_2)) < \hbox{Cor}(Y(\bw_1), Y(\bw_3))$, which is counter-intuitive and not interpretable. As shown in \cref{eg1}, $\hbox{Cor}(Y(\bw_1), Y(\bw_2))$ should be no less than $\hbox{Cor}(Y(\bw_1), Y(\bw_3))$, since $(\bw_1,\bw_2)$ are more similar compared to $(\bw_1, \bw_3)$.
\end{example}

\subsection{The Efficient EzGP (EEzGP) Model}
\label{egagps}

The EzGP model with the covariance function in \cref{eq:corr-fun} has $(2 + p + q + p\sum_{h=1}^{q}m_h)$ parameters.
For data with many qualitative factors, this number can be quite large, which may result in high prediction variance.
In this part, we propose a so-called Efficient EzGP (EEzGP) model for data with many qualitative factors.

The EEzGP model follows the same \cref{eq:additiveGP,eq:cor,eq:agp} as the EzGP, but simplifies the correlation parameter $\theta^{(h)}_{kl_{h}}$ in \cref{eq: cor-1} to $\theta^{(h)}_{l_h}$. It considers $G_{z^{(h)}}(\bx)$ ($h = 1, \ldots, q$) to be a GP with the covariance function:
\begin{align}\label{eq: cor-2}
\phi_{h}((\bx_{i}^\TT, z_{ih})^\TT, (\bx_{j}^\TT, z_{jh})^\TT )
    & = \sigma_h^2 \hbox{exp} \left\{ - \sum\limits_{k=1}^p  \theta^{(h)}_{l_h} (x_{ik} - x_{jk})^2 \right\} I(z_{ih} = z_{jh} \equiv l_h).
\end{align}

Compared to the covariance function in \cref{eq: cor-1} which adopts distinct correlation parameters $\theta^{(h)}_{kl_h}$ to scale each quantitative factor separately,  the covariance function in \cref{eq: cor-2} adopts a single correlation parameter $\theta^{(h)}_{l_h}$ to scale all quantitative factors together.
\rb{As the EEzGP model includes a base GP component $G_0$ where distinct correlation parameters have been used for different quantitative factors, it may not be necessary to scale each quantitative dimension again when considering the coupled quantitative effects in the adjustment part $G_{z^{(h)}}(\bx)$.
Thus, such a simplification may not sacrifice much in model prediction accuracy. Examples in  \cref{ss,re}  will  illustrate this point. When using the EEzGP model, we should always normalize the quantitative factors to $[0,1]$ range. To avoid over-parameterization in \cref{eq: cor-2},  we fix $\theta^{(h)}_{1} = 1$ for the first level in $z^{(h)}$, which can be viewed as a benchmark for the adjustment.}

For two inputs $\bw_i = (\bx_i^\TT,\bz_i^\TT)^\TT$ and $\bw_j = (\bx_j^\TT,\bz_j^\TT)^\TT$, the covariance function $\phi(\bw_i, \bw_j)$ (for any $i,j = 1, \ldots, n$) in the EEzGP model is
\begin{align}\label{cov:mgagp}
\phi(\bw_i, \bw_j) \nonumber & =  \hbox{Cov}(Y(\bw_i), Y(\bw_j)) \nonumber \\
& =  \sigma_{0}^{2}\hbox{exp} \left\{ - \sum_{k=1}^p \theta_{k}  (x_{ik} - x_{jk})^2 \right\} \nonumber \\
&  + \sum_{h=1}^{q}   \sum_{l_h =1}^{m_{h}} I(z_{ih} = z_{jh} \equiv l_h) \sigma_h^2 \hbox{exp} \left\{ - \sum\limits_{k=1}^p  \theta^{(h)}_{l_h} (x_{ik} - x_{jk})^2 \right\}.
\end{align}
\cref{pd,pdc} and \cref{pdcc} in  \cref{GF} also apply to the EEzGP model, since it is a special case of the EzGP. When Latin hypercube designs or space-filling designs are used for quantitative factors, the covariance matrix of observed responses $\hbox{Cov}(\bm y)$ induced by \cref{cov:mgagp} is positive definite.

In the EEzGP model, the number of parameters is $2 + p + \sum_{h=1}^{q}m_h$, which is much smaller than that in the EzGP model.
A rule of thumb for run-size in computer experiments is at least $10(p+q)$, ten times of the dimensions \cite{loeppky2009}.
\rb{Taking $m_1 = \ldots m_q = m$ for illustration, it is easy to show that when the number of levels in qualitative factors $m \leq 10$, the number of parameters in the EEzGP model will be less than $10(p+q)$}.

\subsection{The Localized EzGP (LEzGP) Method}
\label{lezgp}
Note that for the EzGP and EEzGP models, the computational complexity and memory space complexity are $O(n^3)$ and $O(n^2)$, respectively, where $n$ is the size of  training data.
To facilitate the analysis of data with large size $n$, we propose the so-called LEzGP method.
Its key idea is to select a proper subset of training data to fit the EEzGP (or EzGP) model given a target input. For an input $\bw = (\bx^\TT,\bz^\TT)^\TT$ and a target input $\bw^{*} = ((\bx^{*})^\TT,(\bz^{*})^\TT)^\TT$, we denote $N_{\bz}(\bw, \bw^{*})$ to be the number of same levels in their qualitative parts (between $\bz$ and $\bz^{*}$). For example, when $\bz = (1,2,3)^\TT$ and $\bz^{*} = (3,2,1)^\TT$, there is only one same level at the corresponding positions, and thus $N_{\bz}(\bw, \bw^{*}) = 1$. The LEzGP method includes the following three steps:
\begin{enumerate}
\item[Step 1.] Select an appropriate tuning parameter $n_{s}$;

\item[Step 2.] For a chosen target input $\bw^{*}$,  select the training data $\bw_i$ ($i \in \{1,\ldots,n \}$) satisfying $N_{\bz}(\bw_i, \bw^{*}) \ge n_{s}$ to form the key subset, denoted as $K_s$;

\item[Step 3.] Use $K_s$ as the new training set and fit it with the EEzGP (or EzGP) model to make prediction at the target input $\bw^{*}$ in Step 2.
\end{enumerate}
Clearly, the value of $n_{s}$ determines the size of the key subset $K_s$.
It means that the data points in $K_s$ have at least $n_{s}$ number of the same levels as the target input in their qualitative parts. The following example illustrates the first two steps in the LEzGP method.

\begin{example}
Consider a computer experiment with five runs, one quantitative and four qualitative factors. Its design matrix $D$ is shown as below. Suppose that the chosen target input $\bw^{*} = (0.3, 1,2,3,1)^\TT$ and the tuning parameter $n_s=3$. Then, the key subset $K_s$  will only include those runs that have at least 3 same levels as $\bw^{*}$ in their qualitative parts $\bz$.
\renewcommand{\kbldelim}{(}
\renewcommand{\kbrdelim}{)}
\[
  D = \kbordermatrix{
    & x^{(1)} & z^{(1)} & z^{(2)} & z^{(3)} & z^{(4)} \\
    w_1 & 0.1 & 1 & 3 & 3 & 1 \\
    w_2 & 0.3 & 1 & 2 & 1 & 3 \\
    w_3 & 0.2 & 2 & 2 & 3 & 1 \\
    w_4 & 0.9 & 2 & 2 & 3 & 2 \\
    w_5 & 0.5 & 1 & 2 & 3 & 1
  };
  K_s = \kbordermatrix{
    & x^{(1)} & z^{(1)} & z^{(2)} & z^{(3)} & z^{(4)} \\
    w_1 & 0.1 & 1 & 3 & 3 & 1 \\
    w_3 & 0.2 & 2 & 2 & 3 & 1 \\
    w_5 & 0.5 & 1 & 2 & 3 & 1
  }.
\]
\end{example}

One primary rationale of the  LEzGP method is that predictions from the GP model fitted by a relevant subset of data can be more accurate than those from the GP model fitted by the entire training set of large data.
As shown in \cite{huang2016computer}, the predicted response at target input (a.k.a. target response) will be less accurate, if its training set contains certain responses following significantly different GPs compared to that followed by the target response.
In computer experiments with qualitative factors, when an observed input has no or few common qualitative levels as the target input, their responses may follow different GPs.
Thus, for the data with large size, it would be appropriate to exclude such irrelevant data points in predicting the target input.

Generally speaking, in the LEzGP method, a larger $n_s$  chosen in  Step 1 would lead to a smaller key subset $K_s$ in Step 2 and less computation required in Step 3.
A larger $n_s$ and smaller $K_s$ will not necessarily reduce the prediction accuracy.
Choosing a proper value of $n_s$ reduces the computational cost, and could improve the prediction accuracy in certain situations; refer to the \cref{cagp1} in \cref{ss}.
The proper selection of $n_s$ often depends on the budget (e.g. available computing resources), the design matrix of the training data and the target input to be predicted. \rb{In practice, budget is often the key constraint.
\nx{In most cases, one can easily choose an appropriate $n_s$, because a small difference in $n_s$ will lead to a big difference in the run size of $K_s$}; see \cref{cagp1}}.

For a large-size computer experiment with QQ inputs, one general suggestion on $n_s$ is to choose
$q/2 < n_s \leqslant n_{up}$,
where $n_{up}$ is the largest integer such that the size of its key subset $K_s$ is larger than the number of parameters in the model. A rule of thumb for the choice of $n_s$ would be
$q/2 < n_s \leqslant n_{up}^{*}$, where $n_{up}^{*}$ is the integer such that the size of its corresponding key subset is closest to $10(p+q)$ \cite{loeppky2009}. A too small $n_s$ will not be desirable since it will require more computing resources. We suggest to use $n_s > q/2$ here, which can guarantee that every pair of data points in  $K_s$ will have at least one common level in the same qualitative factor.

We would like to note that it is possible for the LEzGP method using other models in its Step 3.
But adopting the EEzGP (or EzGP) model appears to provide better justifications.
The underlying assumption of the LEzGP method is that when two inputs have an increased number of common levels in qualitative factors, these two inputs are more relevant and thus their correlation should increase.
In the covariance function \cref{cov:mgagp} (or \cref{eq:corr-fun}), more positive covariance components due to the same qualitative levels are added when two inputs have more common  levels $l_h$ in $z_h$ ($h = 1, \ldots, q$), which exactly matches the assumption of the LEzGP method.

\subsection{Parameter Estimation}
\label{es}

The EzGP model with the covariance function in \cref{eq:corr-fun}  contains the parameters $\mu$, $\sigma_0^2$, $\theta_{k}^{(0)}$, $\sigma_h^2$ and $\theta_{kl_h}^{(h)}$ where $h = 1, \ldots, q$, $k = 1, \ldots, p$ and $l_h = 1, \ldots, m_h$. Denote vector $\bm{\sigma^2} = (\sigma_0^2, \sigma_1^2, \ldots, \sigma_q^2)^\TT$ and matrix $\bm{\Theta} = (\bm{\theta}^{(0)}, \bm{\Theta}^{(1)}, \ldots, \bm{\Theta}^{(q)})$ where $\bm{\theta}^{(0)} = (\theta^{(0)}_k)_{p \times 1}$ and  $\bm{\Theta}^{(h)} = (\theta_{kl_h}^{(h)})_{p \times m_h}$. Denote the covariance matrix by $ \bm{\Phi} =  \Phi(\bm{\sigma^2}, \bm{\Theta}) = (\phi(\bw_i, \bw_j))_{n \times n}$ which follows the covariance function in  \cref{eq:corr-fun}.
Under the GP model in  \cref{eq:additiveGP} and after dropping some constants, maximizing the log-likelihood function $l(\mu, \bm{\sigma^2}, \bm{\Theta})$ is equivalent to minimizing
$\hbox{log}\vert \bm{\Phi} \vert + (\by-\mu\textbf{1})^\TT\bm{\Phi}^{-1}(\by - \mu\textbf{1})$.
For given $\bm{\sigma^2}$ and $\bm{\Theta}$, the maximum likelihood estimator (MLE) of $\mu$ is $\widehat{\mu} = (\textbf{1}^\TT\bm{\Phi}^{-1}\textbf{1})^{-1}\textbf{1}^\TT\bm{\Phi}^{-1}\textbf{y}$.
Thus we can obtain the profile likelihood for the MLE of $\bm{\sigma^2}$ and $\bm{\Theta}$ :
\begin{equation}
\label{mlep}
[\bm{\sigma^2}, \bm{\Theta}] = \textrm{argmin} \left\{ \log \vert \bm{\Phi} \vert + (\by^\TT \bm{\Phi}^{-1} \by) - (\textbf{1}^\TT\bm{\Phi}^{-1} \textbf{1})^{-1} (\textbf{1}^\TT\bm{\Phi}^{-1} \textbf{y})^2 \right\}.
\end{equation}
\rb{This minimization problem can be solved via some standard global optimization algorithms in R or Matlab, such as genetic algorithms   \cite{ranjan2011computationally, macdonald2013gpfit}.
In this work, we adopt the R package ``rgenoud" \cite{Walter2011} which combines evolutionary search algorithms \cite{lin2015using} with the derivative-based quasi-Newton methods to solve difficult optimization problems.
In particular, we have derived all parameters' analytical gradients to facilitate the computation, which are reported in the \cref{sec:ag}.}

Given parameters $\mu$, $\bm{\sigma^2}$ and $\bm{\Theta}$, the prediction at a new location $\bw^*$ is the condition mean:
\begin{equation}
\label{predy}
\widehat{Y}(\bw^*) = E(Y(\bw^*) \vert y_1, \ldots, y_n) = \widehat{\mu} + \bm{\gamma}^\TT \bm{\Phi}^{-1}(\by - \widehat{\mu} \textbf{1}),
\end{equation}
where $\bm{\gamma}$ is the covariance vector $(\phi(\bw^{*}, \bw_i))_{n \times 1}$ for $i=1,\ldots,n$. As $\widehat{Y}(\bw^*)$ is unbiased, we obtain $\text{MSE}(\widehat{Y}(\bw^*)) = \text{Var}(\widehat{Y}(\bw^*))$, which is expressed as
$
\sum_{i=0}^q\sigma_i^2 - \bm{\gamma}^\TT \bm{\Phi}^{-1}\bm{\gamma} + \frac{(1-\textbf{1}^\TT\bm{\Phi}^{-1}\bm{\gamma})^2}{\textbf{1}^\TT\bm{\Phi}^{-1}\textbf{1}}
$.

For the interpolation property, when $\bw^*$ is the $i^{th}$ observed input $\bw_i$, $\bm{\gamma}^\TT$ is the $i^{th}$ row in $\bm{\Phi}$; thus, $\bm{\gamma}^\TT\bm{\Phi}^{-1}$ is the $i^{th}$ row of $\bm{\Phi} \bm{\Phi}^{-1}$, which is a row vector with its $i^{th}$ entry being 1  and otherwise 0. Therefore, it is straightforward to show $\widehat{Y}(\bw^*) = \widehat{Y}(\bw_i) = y_i$ by \cref{predy}. Similar parameter estimations apply for the EEzGP model which is a special case of the EzGP model. More details on the derivations for the general GP model can be found in \cite{kleijnen2009, rasmussen2006}.

\section{Simulation Study}
\label{ss}
In this section, we use three numerical examples to examine performances of our proposed models.
We measure the performance via the root mean square error (RMSE) for predictions:
$$
\textrm{RMSE} = \sqrt{\frac{1}{n_t}\sum_{i=1}^{n_t}(\widehat{Y}(\bw_i) - Y(\bw_i))^2},
$$
where $n_t$ is the number of data points in the test set, $\hat{Y}(\bw_i)$ and $Y(\bw_i)$ are the predicted and actual responses of the input $\bw_i$ in the test set. In addition, we use the Nash-Sutcliffe efficiency (NSE) \cite{kaufman2011efficient} to describe model's goodness-of-fit, which is defined as
$$
\hbox{NSE} = 1 - \frac{\sum_{i=1}^{n_t}(\hat{Y}(\bw_i) - Y(\bw_i))^2}{\sum_{i=1}^{n_t}(\hat{Y}(\bw_i) - \bar{Y})^2},
$$
where $\bar{Y}$ is the average of the predicted responses. The NSE represents an estimate of the proportion of the response variability explained by the model, which is analogous to
the $R^2$ in linear regression.
Generally speaking, a method with a lower RMSE will a yield higher NSE.

\begin{example}
\label{egplusprod}
Consider a computer experiment with $p=3$ quantitative factors and $q=3$  qualitative factors each having 3 levels, and its computer model has the following form ($\bx = (x_1,x_2,x_3)$):
$$
y = f_i(\bx) \times ( g_j(\bx) + h_k(\bx))
$$
where $i$, $j$, $k$ are the levels for the first, second and third qualitative factors, $0 \leqslant x_i \leqslant 1$ for $i = 1,2,3$, and we list functions $f_i$, $g_j$ and $h_k$ as below :
\begin{align*}
 f_1(\bx) &= x_1 + x_2^2 +x_3^3, \ f_2(\bx) = x_1^2 + x_2 +x_3^3, \\
 f_3(\bx) &= x_1^3 + x_2^2 +x_3, \  g_1(\bx) = \cos(x_1) + \cos(2x_2) +\cos(3x_3),\\
 g_2(\bx) &= \cos(3x_1) + \cos(2x_2) +\cos(x_3), \ g_3(\bx) = \cos(2x_1) + \cos(x_2) +\cos(3x_3),\\
 h_1(\bx) &= \sin(x_1) + \sin(2x_2) + \sin(3x_3), \  h_2(\bx) = \sin(3x_1) + \sin(2x_2) + \sin(x_3),\\
 h_3(\bx) &= \sin(2x_1) + \sin(x_2) + \sin(3x_3).
\end{align*}
\end{example}

\begin{figure}[ht]
\caption{\it The boxplots of RMSEs for different models in \cref{egplusprod}}
  \begin{center}
  \begin{tabular}{c}
    \includegraphics[width=0.55\linewidth]{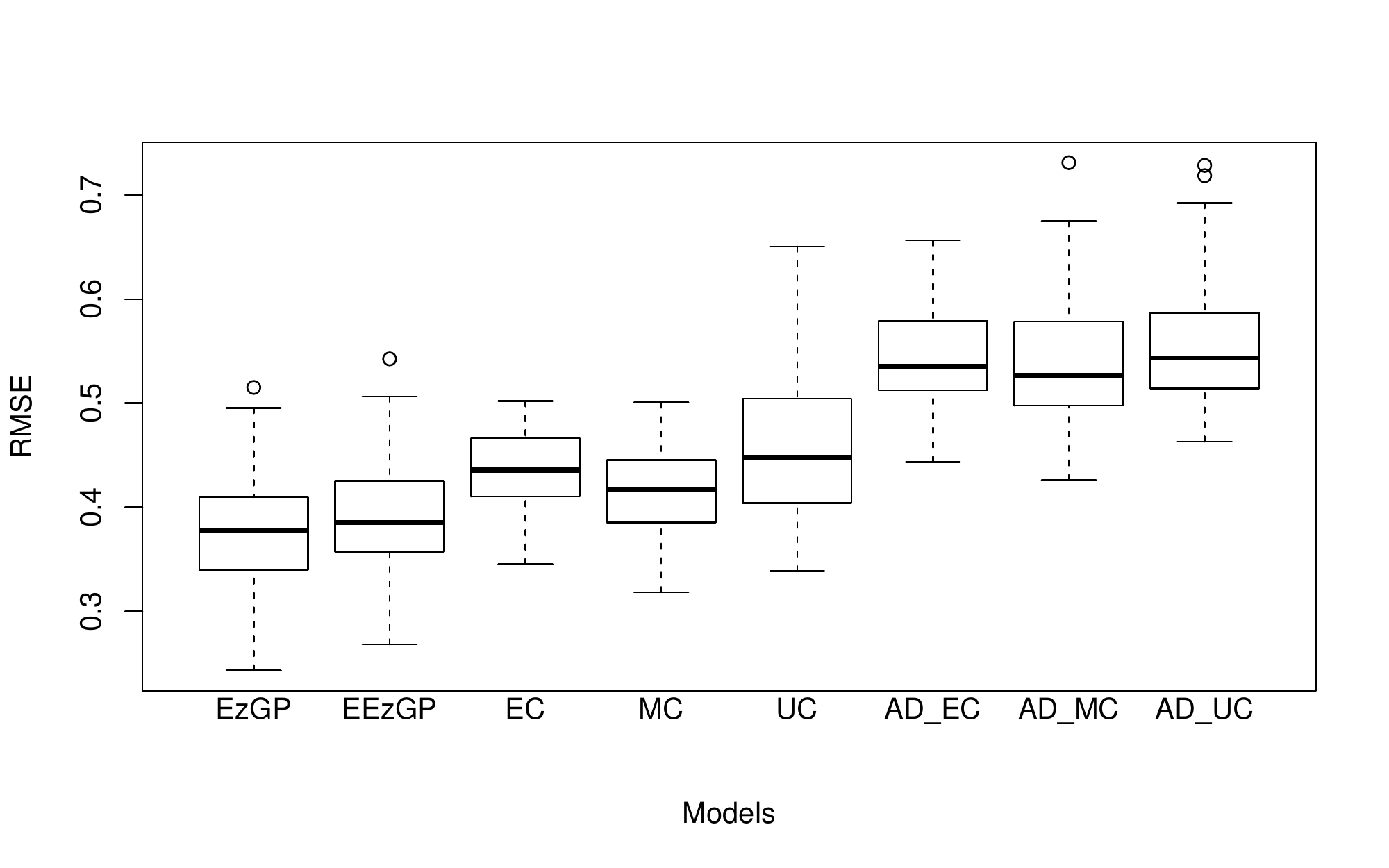}
  \end{tabular}
  \end{center}
    \label{fig:mulplu}
\end{figure}

\rb{In \cref{egplusprod},  the computer model includes both multiplicative and additive structures, which leads to a fair comparison of different multiplicative and additive GP models.}  In  \cref{fig:mulplu}, we show the boxplots of RMSEs for the EzGP, EEzGP, EC, MC, UC, AD\_EC, AD\_MC and AD\_UC models over 50 simulations.
In each simulation, a 81-run design is used where three replicates of a $3^3$ full factorial design are adopted for qualitative factors and a random Latin hypercube design is adopted for quantitative factors. The RMSEs are computed based on a 1215-run test set consisting of 45 replicates of a $3^3$ full factorial design for qualitative factors and a random Latin hypercube design for quantitative factors.

\cref{fig:mulplu} clearly shows that the EzGP and EEzGP models perform better than other models with smaller RMSEs.
Here, the EzGP model performs the best, and it has more parameters than the EEzGP model.
For experiments with a relatively small number of quantitative and qualitative factors, the EzGP model is usually preferred due to its flexibility. The median NSE value for the EzGP model here is as high as 0.92, which is analogous to achieving an $R^2 = 0.92$ in linear regression. Thus, the EzGP model fits the data well.
It should be noted that the multiplicative models EC, MC and UC perform better than the additive models: AD\_EC, AD\_MC and AD\_UC here.
Thus, the success of EzGP and EEzGP methods is not because this simulation setting is in favor of additive models.
\rb{In this example, the computer models are of different expressions for the distinct level combinations of the qualitative factors. The key idea of the proposed models is using the indicator functions appropriately in the GP covaraince function to make the response surface different under the different level combinations of the qualitative factors.
Thus, the superior performances of the EzGP and EEzGP methods here could be explained by using the meaningful additive covariance structures via the  indicator functions.
}

\begin{example}
\label{eghigh}
Consider a computer experiment with $p=9$ quantitative factors and $q=9$  qualitative factors each having 3 levels, and the computer model has the following form:
\begin{align*}
y = & f_{i_1}^{(1)}(x_1,x_2,x_3)g_{j_1}^{(1)}(x_1,x_2,x_3) + f_{i_2}^{(2)}(x_4,x_5,x_6)g_{j_2}^{(2)}(x_4,x_5,x_6) + \\
& f_{i_3}^{(3)}(x_7,x_8,x_9) g_{j_3}^{(3)}(x_7,x_8,x_9) + f_{i_1}^{(1)}(x_7,x_8,x_9)h_{k_1}^{(1)}(x_7,x_8,x_9) + \\
& f_{i_2}^{(2)}(x_4,x_5,x_6)h_{k_2}^{(2)}(x_4,x_5,x_6) + f_{i_3}^{(3)}(x_1,x_2,x_3)h_{k_3}^{(3)}(x_1,x_2,x_3),
\end{align*}
where $0 \leqslant x_i \leqslant 1$ for $i = 1, \ldots, 9$. Here, the nine qualitative factors $z^{(1)}, \ldots, z^{(9)}$ correspond to functions $f^{(1)}$, $f^{(2)}$, $f^{(3)}$, $g^{(1)}$, $g^{(2)}$, $g^{(3)}$, $h^{(1)}$, $h^{(2)}$ and $h^{(3)}$, respectively, and $i_1,i_2,i_3, j_1,j_2,j_3, k_1,k_2,k_3 \in \{1,2,3\}$ are the levels for these nine qualitative factors. We list functions $f$, $g$ and $h$ as below:
\begin{align*}
& f_s^{(l)}(a,b,c) = a^{(r_1+1)} + b^{(r_2+1)} + c^{(r_3+1)}, \\
& g_s^{(l)}(a,b,c) = \cos((r_2+1)a) + \cos((r_1+1)b) + \cos((r_3+1)c), \\
& h_s^{(l)}(a,b,c) = \sin((r_3+1)a) + \sin((r_2+1)b) + \sin((r_1+1)c),
\end{align*}
where parameters $r_1 = s+l+1\ (\text{mod} \ 3)$, $r_2 = s+l+2\ (\text{mod} \ 3)$ and $r_3 = s+l$ (mod 3) for $l=1,2,3$ and $s=1,2,3$.
\end{example}

\begin{figure}[ht]
  \centering
  \caption{\it The boxplots of RMSEs for different models in  \cref{eghigh}}
  \begin{tabular}{c}
    \includegraphics[width=0.55\linewidth]{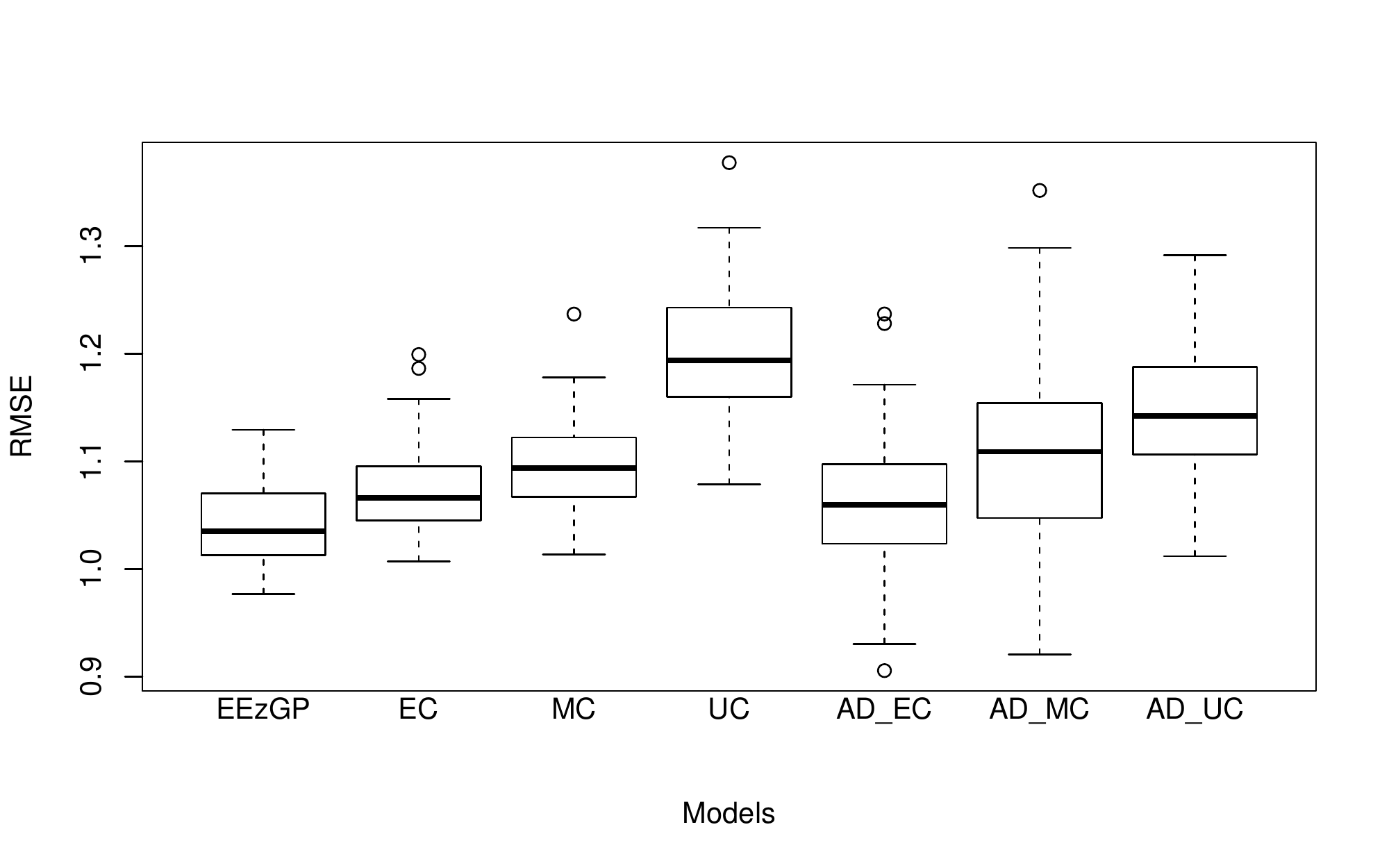}
  \end{tabular}
    \label{large}
\end{figure}

In \cref{eghigh}, the computer experiment has many factors and very complex computer models, which is  suitable to test emulator's prediction power and robustness.
In \cref{large}, we display the boxplots of the RMSEs for each model over 50 simulations. In each simulation, a 243-run design is adopted, where a space-filling 3-level orthogonal array \cite{xiao2018} is used for the qualitative factors and a random Latin hypercube design is used for the quantitative factors. The
RMSEs are computed based on a 1215-run test set consisting of a random 3-level fractional factorial design for the qualitative factors and a random Latin hypercube design for the quantitative factors. From
\cref{large}, we can see that the EEzGP model outperforms all others in terms of the median RMSE.
It is also the most stable model if we look at the worst case scenario. The median NSE for the EEzGP model is 0.76, which is good in practice.
As illustrated in  \cref{egagps}, it is not recommended to use the EzGP model for computer experiments with many factors, and thus we do not compare it here.

\begin{figure}[ht]
  \centering
  \caption{\it The boxplots of RMSEs for the LEzGP in \cref{cagp1} and the EEzGP in \cref{eghigh}}
  \begin{tabular}{c}
    \includegraphics[width=0.50\linewidth]{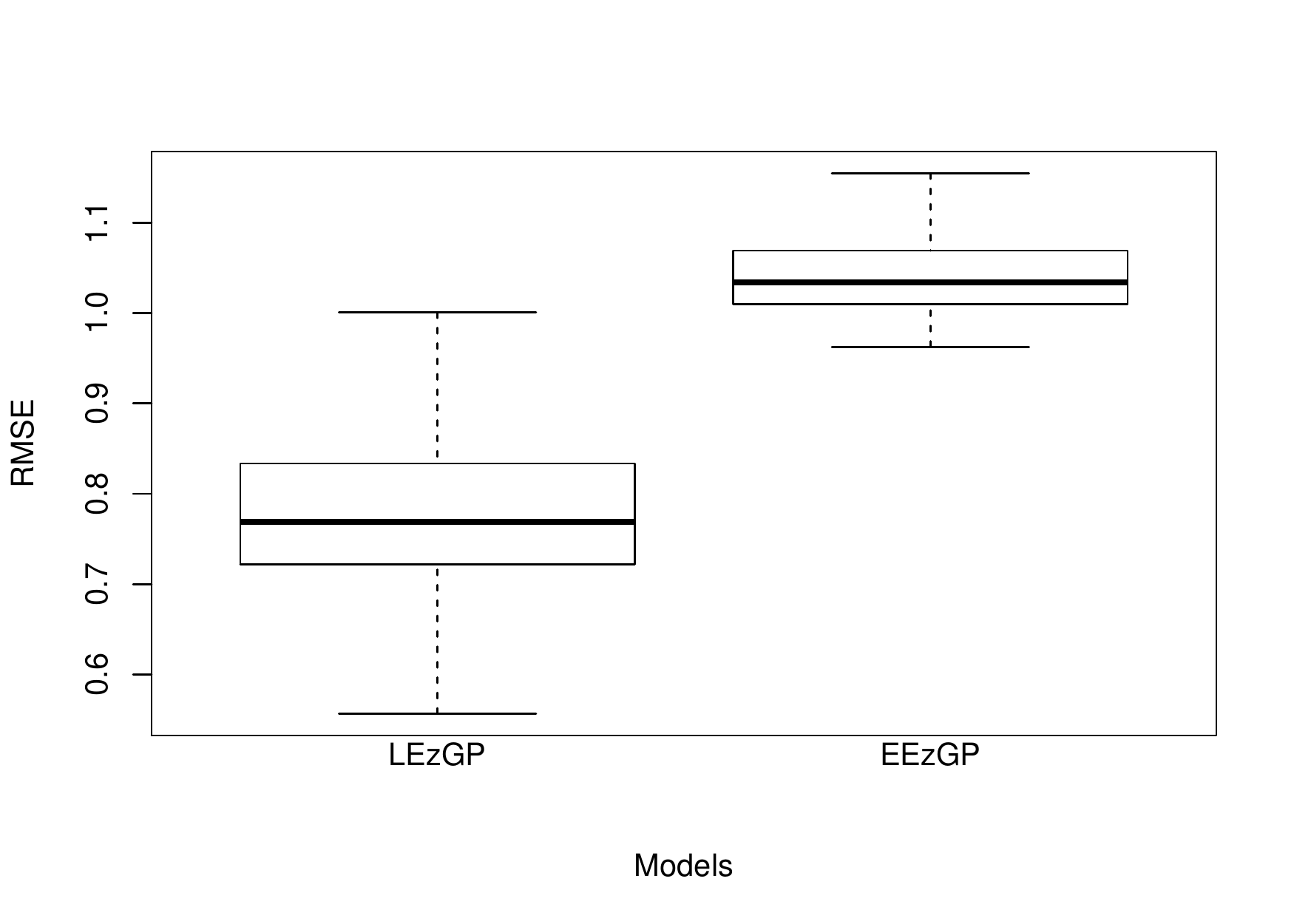}
  \end{tabular}
    \label{cagp}
\end{figure}

\begin{example}
\label{cagp1}
This example is to examine the performance of the proposed LEzGP method. Consider a computer experiment with $n =$ 19,683 runs, $p=9$ quantitative factors and $q=9$  qualitative factors each having 3 levels.
The computer models are the same as those in \cref{eghigh}.
A 19,683-run design is use with a random Latin hypercube design for the quantitative factors and a $3^9$ full factorial for the qualitative factors.
The RMSEs are computed based on a test set consisting of $m = 100$ data points where a random Latin hypercube design used for the quantitative factors and a single random level combination used for the qualitative factors.
We replicate this simulation 50 times and display the boxplots of  RMSEs in \cref{cagp}.
\end{example}

For such a computer experiment with a large run size $n$, it is difficult to directly apply existing GP models (EzGP, EEzGP, EC, MC, UC, AD\_EC, AD\_MC and AD\_UC models).
Thus, the proposed LEzGP method is in a better position for evaluation.
For the LEzGP, it is straightforward to show that the number of data points in the key subset $K_s$ is $m^q[1 - \sum_{i=0}^{n_s-1} {q \choose i}(1/m)^i(1-1/m)^{q-i}]$ with the tuning parameter $n_s$.
We set the tuning parameter $n_s=7$ according to the rule of thumb in \cref{lezgp}, and consequently the LEzGP method selects a $K_s$ of 163 training data from the overall 19,683 ones.
The LEzGP method significantly reduces the computation and memory space required in model estimation.

In \cref{cagp}, we compare the performance of the LEzGP method with that of the EEzGP model in  \cref{eghigh}, since both examples use the same computer model. From \cref{cagp}, the LEzGP method can provide more accurate predictions using only 163 training data, compared with the EEzGP model using 243 training data. The median NSE for the LEzGP method here is 0.87, larger than that of 0.76 for the EEzGP model. Moreover, the success of LEzGP method also provides some justifications on the assumptions of our proposed models: a data point will not contribute much to the prediction of the target input, if it has no same level as the target in their qualitative parts.
\rb{Note that when the tuning parameter $n_s$ is 5, 6, 7, 8 or 9 ($n_s$ must be an integer), the corresponding training set $K_s$ will include 2851, 835, 163, 19 and 1 runs, respectively. Clearly, a $K_s$ with 1 or 19 runs is too small and  a $K_s$ with 2851 runs \nx{can be} too big for the LEzGP model with 38 parameters in this example.
Additional results have shown that using $n_s=6$ (with 835 runs) and $n_s=7$ (with 163 runs) will lead to very similar performances on predictions.
Thus, the rule of thumb $n_s=7$ is preferred, which requires much less computation.}

\section{Real Data Analysis}
\label{re}
\begin{figure}[ht]
  \centering
  \caption{\it An illustration of the full scale embankment structure}
  \begin{tabular}{c}
    \includegraphics[width=0.6\linewidth]{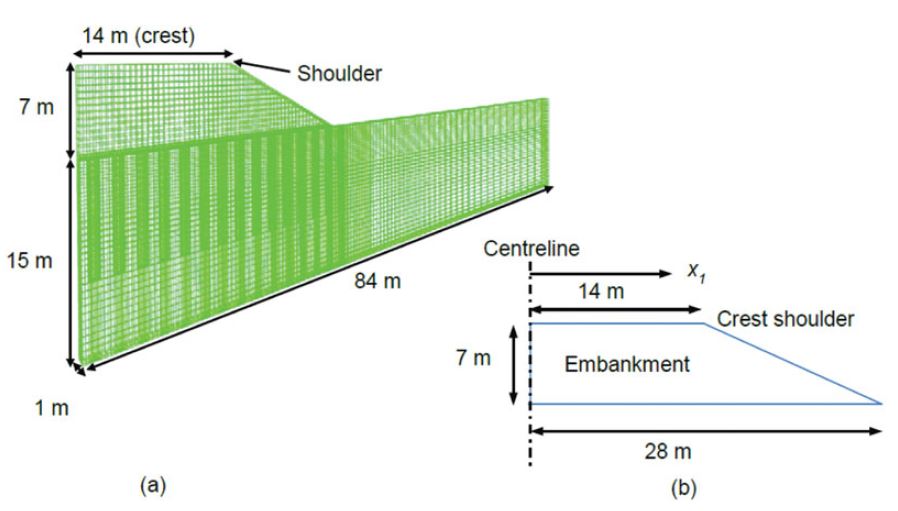}
  \end{tabular}
    \label{realfig}
\end{figure}

In this section, we apply the proposed models to a real computer experiment  with $p=1$ quantitative factor and $q=3$  qualitative factors. A fully 3D coupled finite element model has
been calibrated and verified by successfully modeling the performance of a full-scale embankment constructed on soft soil \cite{rowe2015}. The following \cref{realfig} (source from \cite{deng2017additive}) illustrates the structure of this full scale embankment where sub-figure (a) is the finite element mesh and sub-figure (b) is the schematic view of embankment constructed on foundation soil.
The finite element discretization here  had 36,802 elements and 69,667 nodes.
The average run-time for one case of this size is approximately 9 hours via a 12-noded super-computer at the High Performance Computing Virtual Laboratory (HPCVL). In this computer experiment, the three qualitative
factors are ``embankment construction rate" ($z^{(1)} = 1, 5, 10$ m/month), ``Young’s modulus of columns" ($z^{(2)} = 50, 100, 200 $ MPa), and ``reinforcement stiffness" ($z^{(3)} = 1578, 4800, 8000$ kN/m). The single quantitative factor $x^{(1)}$ is the distance from the embankment shoulder to the embankment center line. The response here is the final embankment crest settlement, which is an important embankment working indicator. The training set of this computer experiment has 261 runs. The quantitative factor $x^{(1)}$ takes
the 29 values uniformly from the interval [0, 14]. For each distinct value of $x^{(1)}$, a 9-run, 3-factor and 3-level fractional factorial design is used for the qualitative factors. The test set has 29 runs where $x^{(1)}$ takes the 29 values uniformly from the interval [0, 14] and $(z^{(1)}, z^{(2)}, z^{(3)}) = (5, 100, 4800)$. Note that such a setting of qualitative
factors is not used in the training set.

\begin{figure}[ht]
  \centering
  \caption{\it The boxplots of log(RMSE) for different models}
  \begin{tabular}{cc}
    \includegraphics[width=0.45\linewidth]{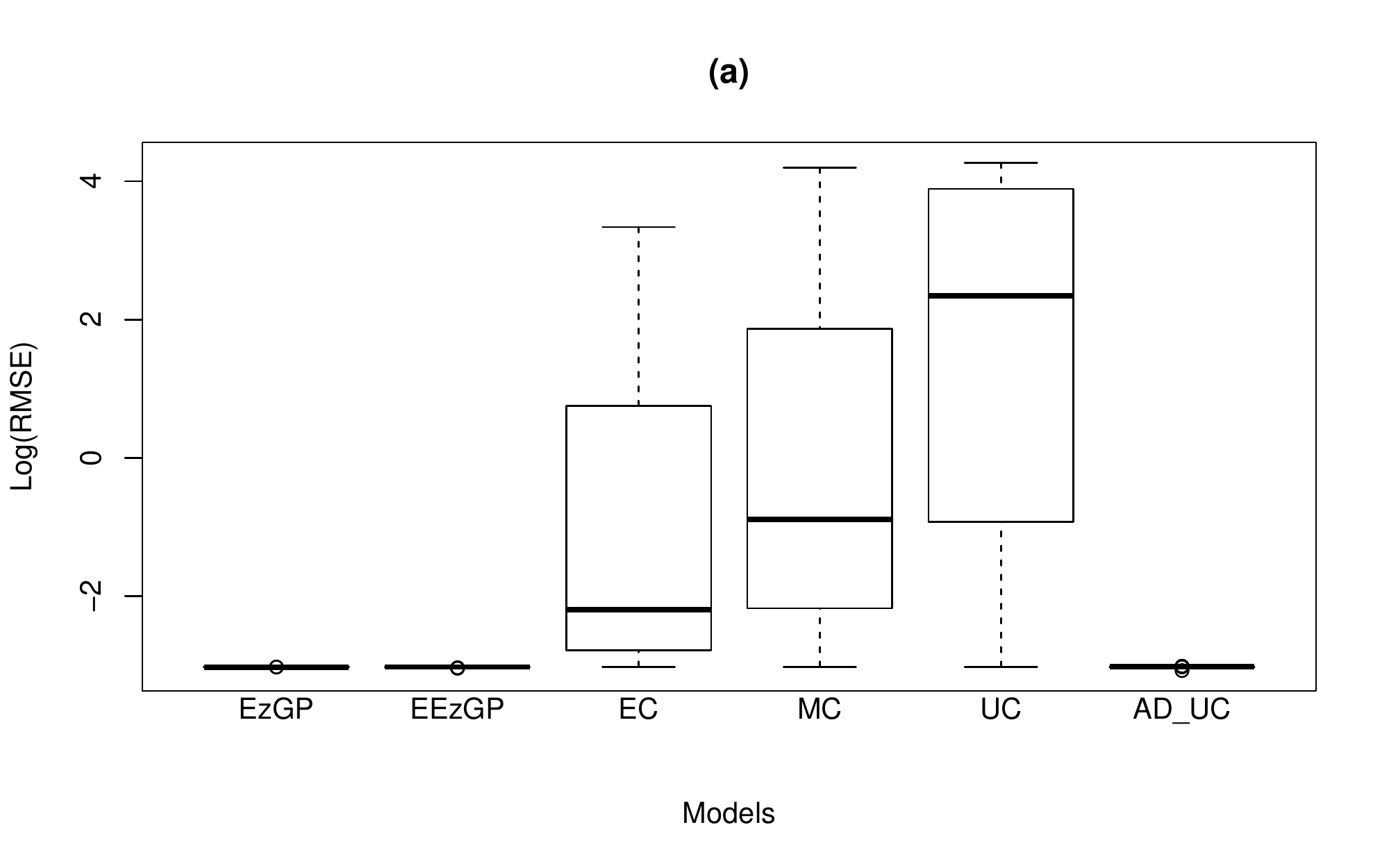} &
    \includegraphics[width=0.45\linewidth]{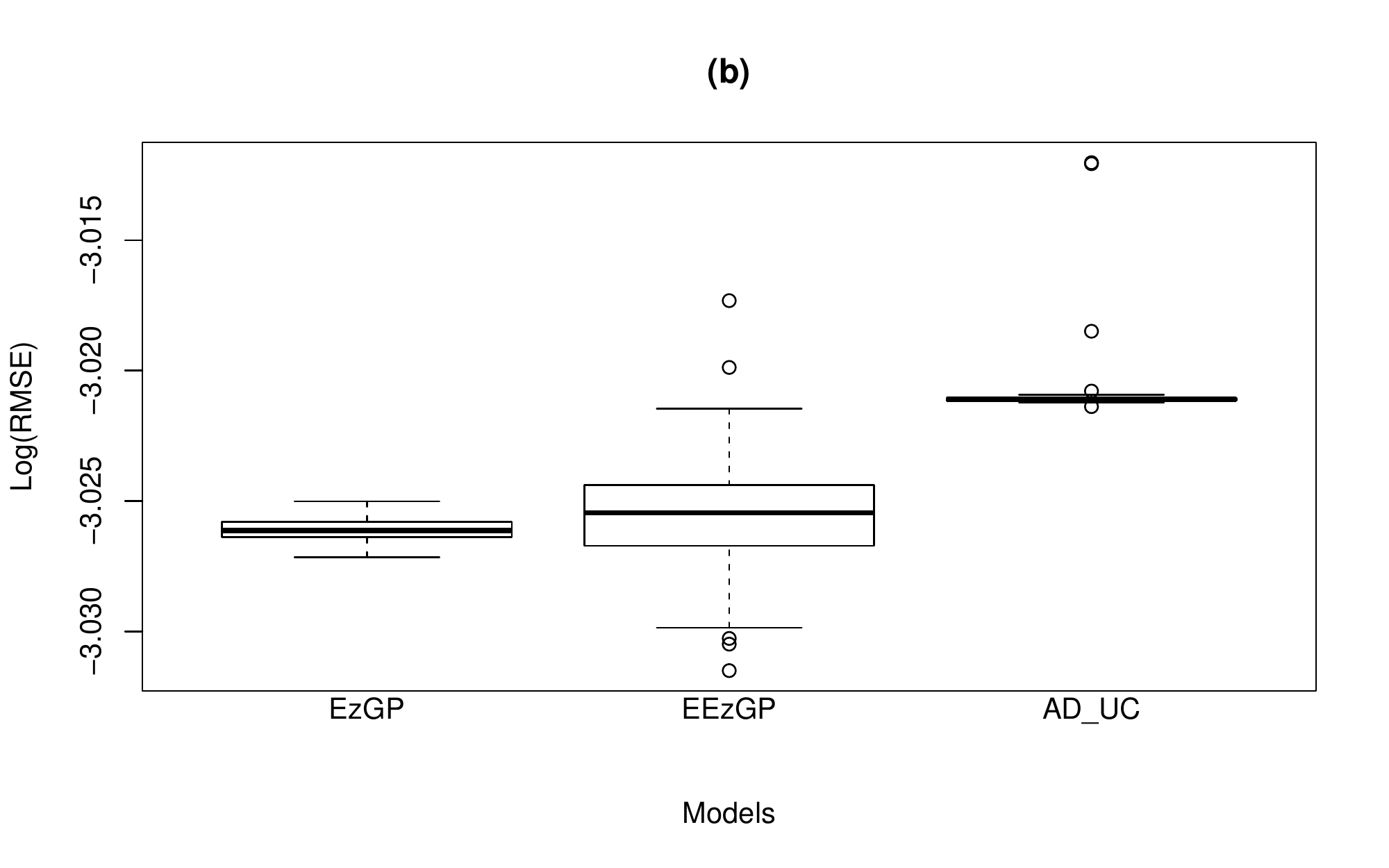}
  \end{tabular}
    \label{xweg}
\end{figure}

To evaluate the proposed methods, we compare the EzGP and EEzGP models with the EC, MC, UC and AD\_UC models as in \cite{deng2017additive}. We repeat each model estimation 100 times as in \cite{deng2017additive}.
\cref{xweg}(a) shows the boxplots of $\hbox{log(RMSE)}$ for different models, and it clearly shows that the EzGP, EEzGP and AD\_UC models perform much better than the EC, MC and UC models. Then, we further compare the EzGP, EEzGP and AD\_UC models in \cref{xweg}(b) and \cref{tab:real}. For the AD\_UC method, we exclude outliers in  \cref{xweg}(b). From the  figure and table, it is clear that the EzGP model performs the best in terms of both mean and median $\hbox{log(RMSE)}$, and it is also the most robust one with the smallest standard deviation.
Note that there is only one quantitative factor and three qualitative factors here.
For cases with only a few factors, the EzGP model is usually preferred due to its flexibility.
The average NSE for the EzGP model is 0.77 which is viewed to be high in practice.
In the EzGP model, the estimate of $\sigma_0^2$ appears to be the largest among those for $\sigma_1^2$, $\sigma_2^2$ and $\sigma_3^2$ in each replication.
This indicates a significant base GP between the output and the quantitative inputs.
It makes practical sense that the distance from the embankment shoulder to the embankment center line has significant impact on the final embankment crest settlement \cite{rowe2015}.
In addition, the estimate of variance parameter $\sigma_1^2$ is larger than that of $\sigma_2^2$ and $\sigma_3^2$, which suggests that the embankment construct rate ($z^{(1)}$) may have stronger impact on the output compared with the other two qualitative factors.

\begin{table}
\center
\caption{Comparison between methods in terms of Log(RMSE)}
\begin{tabular}{lrrr}
\hline \hline
         & Mean     &   Median &    SD    \\
EzGP    & $-3.026$ & $-3.026$ & $0.0005$ \\
EEzGP    & $-3.025$ & $-3.025$ & $0.0040$ \\
AD\_UC  & $-3.021$ & $-3.021$ & $0.0056$ \\
\hline \hline
\end{tabular}
  \label{tab:real}%
\end{table}

\section{Discussion}
\label{cd}
In this work, we propose the EzGP model for computer experiments with both quantitative and qualitative factors, and develop its two useful variants, EEzGP for data with many factors and LEzGP for data with many runs. The proposed models have easy-to-interpret covariance structures and can provide desirable prediction performances.
\rb{Specifically, the proposed models are suitable for handling complex computer experiments with quantitative factors and multiple qualitative factors, where the computer models are very different for the distinct level combinations of the qualitative factors.
Note that the proposed methods quantify the underlying response surfaces of the quantitative factors differently under the different level combinations of the qualitative factors via the additive GP structure. Hence, it is more flexible in terms of quantifying the variance and correlation structure of the quantitative factors compared to \cite{qian2008}, while it could be a bit more restrictive in terms of quantifying the correlation of the qualitative factors due to the use of indicator functions.}

\rb{The current paper focuses on the ``first-order" GP components $G_{z^{(h)}}$ ($h = 1, \ldots q$) in the EzGP framework, which is analogous to the main-effect under the context of GPs. A further research can include the ``second-order" GP components $G_{z^{(h)} z^{(s)}}$ ($h,s = 1, \ldots q$ and $h \neq s$) which can be viewed as the adjustment by the interaction of $z^{(h)}$ and $z^{(s)}$. One can consider its covariance function as $
\phi_{hs}((\bx_{i}^\TT, z_{ih}, z_{is})^\TT, (\bx_{j}^\TT, z_{jh}, z_{js})^\TT ) = I(z_{ih} = z_{jh} \equiv l_h) I(z_{is} = z_{js} \equiv l_s)
\sigma_{hs}^2 \hbox{exp} \{ - \sum\limits_{k=1}^p \theta^{(hs)}_{kl_hls}  (x_{ik} - x_{jk})^2 \}
$. However, such an EzGP (or EEzGP) model may contain too many parameters, and thus may over-fit the data in practice.

Here, we would like to remark that the proposed EzGP framework can provide good interpretations on the importance of qualitative factors via the variance parameters.
To get robust variance parameter estimations and alleviate too complex model structures, one could add a penalty term of the variance parameters to the likelihood function in the proposed models.
Adding a penalty term for GP modeling is used in the literature \cite{hung2011penalized}, and variable screening for computer experiments with QQ inputs can be another topic of future research.

It will be an interesting investigation to further enhance the LEzGP method.
Better strategies of selecting the tuning parameter $n_s$ need to be investigated.}
Other methods in selecting subsets may also be useful in the LEzGP method, e.g. the localization method in \cite{gramacy2015local}.
In addition, one issue of the current LEzGP method is that when there are many different level combinations of qualitative factors in the target inputs, the model estimation can still be computationally cumbersome, if the goal is to predict the whole response surface.
One possible solution is to arrange the target inputs into a few groups according to their level combinations, and then apply a more flexible LEzGP method to each of these groups.

Good experimental designs usually have
significant impacts on both computer and physical experiments \cite{fang2005, santner2003, xiao2018application}. 
For the standard GP models, space-filling designs are usually preferred \cite{xiao2017, lin2015latin, wang2018}.  
The marginally coupled designs were proposed  for computer experiments with QQ inputs \cite{deng2015design}, but their run and factor sizes are not flexible.
Construction of good space-filling designs of flexible sizes for GP models with QQ inputs remains a challenging  problem.

\appendix

\section{Proofs for Theoretical Results}

\begin{proof}[{Proof of \cref{pd}}] By \cref{mezgp}, we have
\begin{equation*}
\hbox{Cov}(\bm y) = ( \phi(\bw_i, \bw_j) )_{n\times n} = \bA_0 + \sum_{h=1}^{q} \sum_{l_h=1}^{m_h} (\bB_{hl_h}\bB_{hl_h}^\TT)  \circ \textbf{A}_{hl_h}.
\end{equation*}
Since $\bA_0 = (\sigma_0^2 R(\bx_i, \bx_j \vert \btheta_0))_{n \times n}$ and $\bA_{hl_h} = (\sigma_h^2 R(\bx_i, \bx_j \vert \btheta_{l_h}^{(h)}))_{n \times n}$ with the Gaussian correlation function $R(\cdot|\btheta)$, it is straightforward that matrices $\bA_0$ and $\bA_{hl_h}$ ($h = 1, \ldots, q$ and $l_h = 1, \ldots, m_h$) are all positive semi-definite \cite{rasmussen2006}. By definition, it is clear that matrices $\bB_{hl_h}\bB_{hl_h}^\TT$ ($h = 1, \ldots, q$ and $l_h = 1, \ldots, m_h$) are also positive semi-definite. According to  Theorem 7.5.3 in \cite{horn2013matrix}, we have
\begin{lemma} (Schur Product Theorem)
\label{sp}
Let $A$ and $B$ be $n \times n$ positive semi-definite matrices, their Schur product $A \circ B$ is positive semi-definite.
\end{lemma}
By \cref{sp}, all $(\bB_{hl_h}\bB_{hl_h}^\TT)  \circ \textbf{A}_{hl_h}$ are positive semi-definite. As the sum of positive semi-definite matrices are still positive semi-definite, the covariance matrix $\hbox{Cov}(\bm y)$ is positive semi-definite.
\end{proof}

\begin{proof}[{Proof of \cref{pdc}}]
In the EzGP model with the covariance function in \cref{eq:corr-fun}, the covariance matrix of $\bm y$ can be written as
$
\hbox{Cov}(\bm y) = ( \phi(\bw_i, \bw_j) )_{n\times n} = \bPhi_0 + \bPhi_1 + \ldots + \bPhi_q,
$
where
$\bPhi_0 = (\sigma_{0}^{2}R(\bx_{i},\bx_{j}|\btheta_{0}))_{n \times n}$ with the Gaussian correlation function $R(\cdot|\btheta)$ is positive semi-definite. Let $\bz^{(h)}$ be the $n \times 1$ column vector of the $h^{th}$ qualitative factor. There exists an $n \times n$ permutation matrix $\textbf{P}$ such that $\textbf{P}\bz^{(h)}$ is the sorted vector $(1,\ldots,1,2,\ldots,2, m_h, \ldots, m_h)^\TT$. Let $\bPhi_h^\TT$ be the covariance matrix corresponding to the permuted data by $\textbf{P}$, and $\bPhi_h^\TT = \textbf{P} \bPhi_h \textbf{P}^\TT$.  By \cref{eq: cor-1}, we have $\phi_h((\bx_{i}, z_{ih}), (\bx_{j}, z_{jh}) | \bTheta^{(h)}) = 0$ when $z_{ih} \neq z_{jh}$, and thus $\bPhi_h^\TT$ is block diagonal where
\[ \bPhi_h^\TT =
\begin{pmatrix}
\textbf{B}^{(h)}_1\\
&\!\!\textbf{B}^{(h)}_2\\
&&\ddots\\
&&&\textbf{B}^{(h)}_{m_h}
\end{pmatrix}.
\]
For $l = 1, \ldots, m_h$, let $n_l$ be the number of level $l$ in $\bz^{(h)}$; $\textbf{B}_l^{(h)} = (\sigma_{h}^{2} R(\bx_{i},\bx_{j}|\btheta_{l}^{(h)}))_{n_l \times n_l}$
 and
$R(\bx_{i},\bx_{j}|\btheta_{l}^{(h)})= \hbox{exp} \{ - \sum_{k=1}^p \theta^{(h)}_{kl}  (x_{ik} - x_{jk})^2\}$ which is a Gaussian correlation function. Thus, $\textbf{B}_l^{(h)}$ is positive semi-definite for $l = 1, \ldots, m_h$, and then $\bPhi_h^\TT$ is positive semi-definite. Since $\bPhi_h^\TT = \textbf{P} \bPhi_h \textbf{P}^\TT$, it is straightforward to prove that $\bPhi_h$ is also positive semi-definite.
If there exists an $h$ ($h = 1, \ldots, q$) such that $\bx_i \neq \bx_j$ whenever $z_{ih} = z_{jh}$, all diagonal matrices $\textbf{B}_l^{(h)}$ ($l=1, \ldots, m_h$) in $\bPhi_h^\TT$ are positive definite, and thus $\bPhi_h^\TT$ and then $\bPhi_h$ are positive definite. Finally, we have $\bPhi = \bPhi_0 + \bPhi_1 + \ldots \bPhi_q$ is  positive definite.
\end{proof}

\section{Expressions of Likelihoods and Analytical Gradients}
\label{sec:ag}
Under notations in \cref{gagps}, the likelihood is
\begin{equation}
\label{lk}
L = \frac{1}{(2\pi)^{n/2}\vert \bPhi\vert^{1/2} } \text{exp}\left( -\frac{1}{2} (\by - \mu \bone_n)^\TT \bPhi^{-1} (\by - \mu \bone_n) \right)
\end{equation}
where the covariance matrix $ \bm{\Phi}$ depends on the parameters $\bm{\sigma^2}$ and $\bm{\Theta}$.
Writing the first order conditions results in analytical expressions for $\mu$ as a function of $\bm{\sigma^2}$ and $\bm{\Theta}$:
$$
\widehat{\mu} = (\textbf{1}^\TT\bm{\Phi}^{-1}\textbf{1})^{-1}\textbf{1}^\TT\bm{\Phi}^{-1}\textbf{y}.
$$
Therefore maximizing the likelihood in \cref{lk} is equivalent to maximizing the ``concentrated" log-likelihood obtained by plugging in the expression of $\mu$ (over the parameters $\bm{\sigma^2}$ and $\bm{\Theta}$):
\begin{equation*}
 -2l(\bm{\sigma^2}, \bm{\Theta}) = n\log(2\pi) + \log \vert \bm{\Phi} \vert + (\by - \widehat{\mu})^\TT \bm{\Phi}^{-1}(\by - \widehat{\mu}).
\end{equation*}
For any parameters inside $\bm{\Phi}$, the expression of the analytical gradient given $\widehat{\mu}$ is:
$$
-2\frac{\partial l}{\partial \bullet} = tr(\bm{\Phi}^{-1}\frac{\partial \bm{\Phi}}{\partial \bullet}) - (\by - \widehat{\mu})^\TT \bm{\Phi}^{-1} \frac{\partial \bm{\Phi}}{\partial \bullet} \bm{\Phi}^{-1}(\by - \widehat{\mu}).
$$
Specifically, for the EzGP model with the covariance function in \cref{eq:corr-fun}, for any $i,j = 1, \ldots, n$, we have:
$$
\frac{\partial \bm{\Phi}}{\partial \sigma_0^2} = \left( \hbox{exp} \{ - \sum_{k=1}^p \theta_{k}^{(0)}  (x_{ik} - x_{jk})^2\} \right)_{n \times n},
$$
$$
\frac{\partial \bm{\Phi}}{\partial \sigma_h^2} = \left( \sum_{l_{h}=1}^{m_{h}} I(z_{ih} = z_{jh} \equiv l_{h}) \hbox{exp} \{ - \sum_{k=1}^p \theta^{(h)}_{kl_{h}}  (x_{ik} - x_{jk})^2 \} \right)_{n \times n}, \text{for } l = 1, \ldots, q,
$$
$$
\frac{\partial \bm{\Phi}}{\partial \theta_{k^*}^{(0)}} = \left(
-\sigma_{0}^{2}(x_{ik^*} - x_{jk^*})^2\hbox{exp} \{ - \sum_{k=1}^p \theta_{k}^{(0)}  (x_{ik} - x_{jk})^2\}
 \right)_{n \times n},
$$

$$
\frac{\partial \bm{\Phi}}{\partial \theta_{k^*l_h^*}^{(h^*)}} = \left(
-\sigma_{h^*}^{2} (x_{ik^*} - x_{jk^*})^2 \hbox{exp} \{ - \sum_{k=1}^p \theta^{(h^*)}_{kl_{h}^*}  (x_{ik} - x_{jk})^2 \} I(z_{ih^*} = z_{jh^*} \equiv l_{h}^*)
 \right)_{n \times n}.
$$
The above expressions of likelihoods and analytical gradients also apply to the EEzGP model, since it is a special case of the EzGP model.


\bibliographystyle{siamplain}
\bibliography{references}
\end{document}